\documentclass[aps,prc,twocolumn,floatfix,showpacs,a4paper,nofootinbib,amsmath,amssymb]{revtex4-2}
\usepackage{graphicx} 
\usepackage{dcolumn}  
\usepackage{bm}       
\usepackage[dvipsnames]{xcolor}
\usepackage{float}    
\usepackage{xspace}       

\newcommand{\be}{\begin{equation}}
\newcommand{\ee}{\end{equation}}
\newcommand{\ba}{\begin{eqnarray}}
\newcommand{\ea}{\end{eqnarray}}
\newcommand\en{\ensuremath{\sqrt{s}}\xspace}
\newcommand\enn{\ensuremath{\sqrt{s_\mathrm{NN}}}\xspace}
\newcommand{\jpsi}{\ensuremath{\mathrm{J}/\psi}\xspace}

\newcommand\pp{\ensuremath{\textrm{pp}}\xspace}

\newcommand\pbpb{\ensuremath{\textrm{Pb--Pb}}\xspace}
\newcommand\xexe{\ensuremath{\textrm{Xe--Xe}}\xspace}
\newcommand\auau{\ensuremath{\textrm{Au--Au}}\xspace}
\newcommand\mean[1]{\ensuremath{\langle#1\rangle}\xspace}
\newcommand\bett{\beta_{\textrm{T}}}
\newcommand\bets{\beta_{\textrm{s}}}

\newcommand\gamt{\gamma_{\textrm{T}}}

\newcommand{\etz}        {\ensuremath{\eta}}
\newcommand{\rhz}        {\ensuremath{\rho^0}\xspace}
\newcommand{\lmbc}       {$\Lambda_c^+$}
\newcommand{\almbc}      {~$\overline{\Lambda^-\mkern-12mu}_c$~}
\newcommand{\lmb}        {\ensuremath{\Lambda}}
\newcommand{\almb}       {\ensuremath{\overline{\Lambda}}}
\newcommand{\lmbr}        {\ensuremath{\Lambda(1520)}~}

\newcommand{\Om}{$\Omega^-$}
\newcommand{\Mo}{$\overline{\Omega^+\mkern-12mu}$~}
\newcommand{\X}{$\Xi^-$}
\newcommand{\Ix}{$\overline{\Xi^+\mkern-12mu}$~}
\newcommand{\Sirp}{$\Sigma(1385)^+$\xspace}
\newcommand{\Sirm}{$\Sigma(1385)^-$\xspace}

\newcommand{\Sirs}{$\Sigma(1385)^{\pm}$\xspace}

\newcommand\vv{\ensuremath{v_{2}}\xspace}

\newcommand\ecc{\ensuremath{\varepsilon_{2}}\xspace}
\newcommand\pt{\ensuremath{p_{\textrm{T}}}\xspace}
\newcommand\mpt{\mean{\pt}}
\newcommand\mt{\ensuremath{m_{\textrm{T}}}\xspace}
\newcommand\td{\ensuremath{\textrm{d}}}
\newcommand\dEta{\ensuremath{\td \eta}\xspace}

\newcommand\dNchdEta{\ensuremath{\td N_{\rm ch}/\dEta}}
\newcommand\dNchdEtapp{\ensuremath{\td N^{\rm pp}_{\rm ch}/\dEta}}

\newcommand\mch{\ensuremath{M_{\rm ch}}\xspace}
\newcommand\mchpp{\ensuremath{M^{\rm pp}_{\rm ch}}\xspace}
\newcommand\mdNde{\mean{\dNchdEta}_{\eta=0}}
\newcommand\mdNdepp{\mean{\dNchdEtapp}_{\eta=0}}

\newcommand{\Npart}        {\ensuremath{N_\mathrm{part}}}

\newcommand{\pizero}          {\ensuremath{\pi^{0}}\xspace}
\newcommand{\pip}          {\ensuremath{\pi^{+}}}
\newcommand{\pim}          {\ensuremath{\pi^{-}}}
\newcommand{\pipmns}          {\ensuremath{\pi^{\pm}}}
\newcommand{\pipm}          {\ensuremath{\pi^{\pm}}\xspace}
\newcommand{\kap}          {\ensuremath{\mathrm{K}^{+}}}
\newcommand{\kam}          {\ensuremath{\mathrm{K}^{-}}\xspace}
\newcommand{\kapm}          {\ensuremath{\mathrm{K}^{\pm}}}
\newcommand{\kzero}        {\ensuremath{{\mathrm K}^{0}_\mathrm {S}}}
\newcommand{\p}              {\ensuremath{\mathrm{p}}}
\newcommand{\pbar}         {\ensuremath{\mathrm{\overline{p}}}}
\newcommand{\kstar}        {\ensuremath{{\mathrm K}^{*0}}}
\newcommand{\akstar}      {\ensuremath{\overline{\mathrm K^{*0}\mkern-15mu}}~}
\newcommand{\phm}        {\ensuremath{\phi (1020)}}
\newcommand{\Dpm}          {\ensuremath{\mathrm{D}^{\pm}}\xspace}
\newcommand{\Dzero}        {\ensuremath{{\mathrm D}^{0}}}

\newcommand{\gevc}         {\textrm {GeV}/{\it c}\xspace}

\begin{document}

\title{Blast-wave-Tsallis-power model for $p_{\rm T}$-spectra and elliptic flow $v_2$ of hadrons\\ in collisions of identical nuclei at energies available at the Large Hadron Collider}

\author{Smbat Grigoryan}
\email{Smbat.Grigoryan@cern.ch}
\affiliation{Joint Institute for Nuclear Research, 141980 Dubna, Russia}
\affiliation{A.I.Alikhanyan National Science Laboratory (YerPhI) Foundation, 0036 Yerevan, Armenia}

\date{\today}
\begin{abstract}
A generalization of the phenomenological blast-wave-Tsallis-power model, recently proposed by the author for the hadrons transverse momentum ($p_\textrm{T}$) spectra measured at the LHC, is developed to also describe the hadrons $p_\textrm{T}$-differential elliptic flow coefficients $v_2$ in identical nuclei collisions of different centralities. The model describes well the available data on $p_\textrm{T}$-spectra and $v_2$ for any $p_\textrm{T}$ of various particles, from pions to charmonia, in Pb--Pb at $\sqrt{s_\mathrm{NN}}=$ 2.76 and 5.02~TeV, and in Xe--Xe at $\sqrt{s_\mathrm{NN}}=$ 5.44~TeV. Also, predictions for OO collisions at $\sqrt{s_\mathrm{NN}}=$ 5.36~TeV are given. While the model is mainly targeting the LHC energies, it also works at much lower RHIC energies.
\end{abstract}

\pacs{24.10.Pa, 13.85.Ni, 12.38.Mh, 25.75.-q} 

\maketitle

\section{\label{sec1}Introduction}

Recently, the phenomenological blast-wave-Tsallis-power model (BWTPM) was
proposed, which describes quite accurately the midrapidity \pt-spectra of various hadrons in the whole \pt range measured in \pp and \pbpb collisions at the LHC~\cite{G21}. These \pt-spectra are integrated over the hadron azimuthal angle $\varphi$. They depend strongly on the collision centrality and center-of-mass energy per nucleon-nucleon pair \enn. BWTPM has three terms including a standard blast-wave model (BWM) term~\cite{BW} to account for the hydrodynamic radial flow and kinetic freeze-out effects, a term in the Tsallis (Ts) distribution form~\cite{Tsallis,G17}, responsible for the contribution from the resonance decays, and a power-law (Po) term describing the QCD hard processes. As shown in Ref.~\cite{G21}, the model parameters depend generally on the collision system (\pp or \pbpb), but some of them are independent of the \enn or the particle type or the collision centrality.

In the present paper, a generalization of the BWTPM is developed to describe, in addition to the \pt-spectra, also the midrapidity elliptic flow coefficients \vv in identical nuclei (AA) collisions. The \vv characterizes the azimuthal asymmetry of hadron momentum distribution and depends on hadron type, \pt, collision system, and centrality. The asymmetry is driven mainly by an almond shape (in the plane transverse to the collision axis) of the hadronic medium (quark-gluon plasma) created in non-central collisions. Due to this shape, the in-medium pressure gradient and radial flow velocity become anisotropic resulting in the anisotropy of the hadron momentum spectra~\cite{Oll1}. Following Ref.~\cite{BW1}, I use the generalized BWM with a transversally anisotropic freeze-out hyper-surface of elliptical shape, instead of the isotropic circular one considered in Refs.~\cite{G21,BW}. For the hadronic medium radial flow velocity I assume similar azimuthal dependence as in Refs.~\cite{BW1,BW2}. Also, the other two therms of BWTPM are modified by including a simple dependence on $\varphi$. Having redefined so the model, the hadron \vv can be determined as the second coefficient in the Fourier series of its invariant yield $\td^3N / (\pt\td\pt\td y\td\varphi)$~\cite{BW1,BW2,Vol1,Vol2,HL99}
\be  \label{eq:v2N3}
\hskip-2mm\vv = \int_0^{2\pi} d\varphi \cos(2\varphi)\frac{\td^3N}{\td\pt\td y\td\varphi}~\bigg/ \int_0^{2\pi} d\varphi \frac{\td^3N}{\td\pt\td y\td\varphi}~,
\ee
where $y$ is the hadron rapidity. The \vv contribution dominates in this series for non-central collisions.

The next objective of the present paper is to reduce strongly the number of the BWTPM parameters by parametrizing their dependence on the collision system, centrality, energy, and on the particle type and \pt using simple formulae. As a measure of collision centrality, the corresponding charged-particle multiplicity density at midrapidity ($\eta$ is pseudo-rapidity)
\be  \label{eq:mch0}
\mch \equiv \mdNde
\ee
is used which serves as a good scaling variable for the comparison of hadron production mechanisms in different collision systems (see, e.g., the ALICE review paper~\cite{alic1}). These parametrizations  increase significantly the predictive power of the model. Furthermore, I propose another simplification of the BWTPM. While in Ref.~\cite{G21} it had two different functional forms for $\pt \leq 40$~GeV and $\pt > 40$~GeV, it now has the same form for any \pt.

The values of the BWTPM parameters are obtained then from a simultaneous fit of the combined data on different hadron \pt-spectra and \vv measured mostly by the ALICE experiment in \pbpb and \xexe collisions at the LHC. In addition to the hadrons considered in Ref.~\cite{G21} (\pipm, \pizero, \kapm, \kzero, \p, \pbar,  \kstar, \akstar, $\phi$, \lmb, \almb, \X, \Ix, \Om, \Mo, \Dzero, \Dpm, \jpsi, and unidentified charged-particles), the \etz, \rhz mesons and \lmbc,\almbc baryons are included too.
It is assumed that a particle and its antiparticle have the same \pt-spectrum and \vv. While in the LHC energy domain, which is the main target of present study, this assumption is perfectly valid, it could be violated at much lower energies. To test the model at such energies, I included in the simultaneous fit also some limited data on \pt-spectra and \vv measured at the RHIC.

Note that the previous similar simultaneous fits of the LHC data on hadrons \pt-spectra and \vv were done mainly at low \pt ($\pt\lesssim 3$~\gevc)~\cite{BW2,alic1,alic2} using various generalizations of the BWM. The present model, due to its additional power-law terms, is able to describe much larger dataset in the whole available \pt-region. 

The paper is organized as follows: Sec.~\ref{sec2} gives details of the generalized BWTPM. Sec.~\ref{sec3} presents a parametrization for the \mch data measured in various AA collisions. Sec.~\ref{sec4} describes parametrizations for the model parameters dependence on the AA collision characteristics and the particle type and \pt. In Sec.~\ref{sec5} the model free parameters are determined from the simultaneous fit of all the used data. In Sec.~\ref{sec6} the results and predictions are discussed. Conclusions are given in the last section.

\section{\label{sec2}Description of BWTPM}

The present model consists of three terms that generalize the similar terms of Ref.~\cite{G21}. The first (BW) term corresponds to the generalized BWM~\cite{BW1} and two others (named Ts and Po as in Ref.~\cite{G21}) include additionally the second Fourier coefficients $v_{2t}(\pt)$ and $v_{2p}(\pt)$, respectively, to account for the elliptic flow. Thus, for a particle $\varphi$-dependent \pt-spectrum in AA collisions at midrapidity, where it can be considered independent of the rapidity $y$, I propose the following model:
\be \label{eq:d3N}
\frac{\td^3 N}{\td\pt\td y\td\varphi} \equiv  F_\varphi \,,
\ee
\vspace{-1.0em}
\begin{align} \label{eq:Fv}
& F_\varphi(\pt) = \frac{g}{2\pi^3}\frac{f_N}{f_S}\pt\mt\left[f_{BW}V\int_0^1 \hat{r} d\hat{r}
\int_0^{2\pi} \frac{d\phi}{2\pi}\right. \nonumber\\
&\left. K_1(\gamt\frac{\mt}{T})
\exp{(\gamt \bett \frac{\pt}{T}\cos(\varphi-\phi_b))}\right.\nonumber\\
& + \left.f_1 V_1[1+2v_{2t}(\pt)\cos(2\varphi)](1+ \frac{\mt}{c_1 e_1 n})^{-c_1 n}\right.\nonumber\\
& + \left.f_2 V_2[1+2v_{2p}(\pt)\cos(2\varphi)](c_2^2+\frac{\pt^2}{e_2^2})^{-\frac{n}{2}}\right], 
\end{align}
where N is the particle yield per collision event, $g=2J+1$ is its spin degeneracy factor, $\mt=\sqrt{m^2+\pt^2}$ is its transverse mass, $\gamt=1/\sqrt{1-\bett^2}$, $\bett$ is the hadronic medium radial (transverse) flow velocity at the kinetic freeze-out
\be  \label{eq:bt}
\bett = \tanh{\rho},\quad \rho=[\rho_0+\rho_2\cos(2\phi_b)]\hat{r}^k,
\ee
$\rho$ is the flow rapidity with the second order anisotropy $\rho_2$ and profile exponent $k$, $V=\pi R_x R_y\tau_f$ is the volume of this medium with transverse sizes $R_x$ and $R_y$, $\tau_f$ and $T$ are its proper life-time and temperature, and $K_1$ is a modified Bessel function of the second kind. Note that $\tanh{\rho_0}$ is equal to $\bets$ used in Ref.~\cite{G21}. The global normalization parameter $f_N$ is close to unity at LHC energies but could depend on the normalization conditions of the used data obtained in different experiments. Its energy dependence is significant at lower RHIC energies and will be specified in Sec.~\ref{sec4}. Parameter $f_S$ is equal to unity (as in Ref~\cite{G21}) for ordinary hadrons consisting of up and down quarks only. But for hadrons containing strange or heavier quarks it is greater than unity and quantifies the suppression of their yields depending on the hadron type, \pt, and collision centrality. It describes the observed enhancement of the relative yields of such hadrons with respect to ordinary ones when going from peripheral to central collisions (i.e., from small to large \mch)~\cite{alic1}. The explicit form of $f_S$ is specified in Sec.~\ref{sec4}.

The azimuthal dependence of $\bett$ in Eq.~(\ref{eq:bt}) is similar to the one in Refs.~\cite{BW1,BW2} while its dependence on radial variable $\hat{r}$ is defined for arbitrary $k$~\cite{G21,BW} which varies strongly with the collision centrality ($k=1$ in Refs.~\cite{BW1,BW2}). The meaning of azimuthal angles $\phi_b$ and $\phi$ and their interrelation
\be
\tan\phi_b = \sqrt{r_{xy}}\tan\phi, ~~r_{xy}\equiv R_x^2/R_y^2
\ee
are explained in Refs.~\cite{BW1,BW2}. In the present model, three parameters $R_x, R_y$, and $\tau_f$ enter into two independent combinations via parameters $V$ and $r_{xy}\leq 1$. 

The important quantities for the elliptic flow \vv are $\rho_2, r_{xy}$~\cite{BW1,BW2}, $k$, and functions $v_{2t}(\pt)$ and $v_{2p}(\pt)$ in Eq.~(\ref{eq:Fv}), though the $T$ and $\rho_0$ also influence the \pt-dependence of \vv~\cite{BW1}. From hydrodynamic models and experimental data it follows that there is an approximately
linear relation between \vv and the eccentricity \ecc of the collision geometry~\cite{Vol3,Oll2,Heinz1,alic3}, which can be defined in the present model as
\be  \label{eq:ecc2}
\ecc=(1-r_{xy})/(1+r_{xy}).
\ee
Moreover, in the generalized BWM~\cite{BW1} with $\rho_2=0$ one gets $\vv=\ecc/2$. 
I have checked by fitting the \vv data that parameter $\rho_2$ as well as the functions $v_{2t}(\pt)$ and $v_{2p}(\pt)$ have centrality dependence similar to $\ecc$. For the model's simplicity, I assume that they are proportional to $\ecc$. Their explicit forms will be specified in Sec.~\ref{sec4}. All other parameters in Eq.~(\ref{eq:Fv}) have the same meaning as in Ref.~\cite{G21}. Note that parameters $f_{BW}, f_1, f_2, c_1$, and $c_2$ are the same for any AA collision system and depend only on the particle type. 

Now, integration of Eq.~(\ref{eq:d3N}) by $\varphi$ gives

\vspace{-1.0em}

\be \label{eq:d2N}
\frac{\td^2 N}{\td\pt\td y} \equiv  F\,, 
\ee
\vspace{-1.0em}
\begin{align} \label{eq:F}
& F(\pt) = \frac{g}{\pi^2}\frac{f_N}{f_S}\pt\mt \left[f_{BW} V \int_0^1 \hat{r} d\hat{r} \int_0^{2\pi} \frac{d\phi}{2\pi} \right. \nonumber\\
&\left. K_1(\gamt\frac{\mt}{T})I_0(\gamt \bett \frac{\pt}{T})
+ f_1 V_1(1+ \frac{\mt}{c_1 e_1 n})^{-c_1 n}\right.\nonumber\\
& + \left.
f_2 V_2(c_2^2 +\frac{\pt^2}{e_2^2})^{-\frac{n}{2}}\right].
\end{align}
Performing a similar integration in Eq.~(\ref{eq:v2N3}), one finds (see also Ref.~\cite{BW2})

\begin{align} \label{eq:v2}
& v_2(\pt) = \frac{g}{\pi^2}\frac{f_N\pt\mt}{f_S F(\pt)} \left[f_{BW}V
\int_0^1 \hat{r} d\hat{r} \int_0^{2\pi} \frac{d\phi}{2\pi} \cos(2\phi_b)\right. \nonumber\\
&\left. K_1(\gamt\frac{\mt}{T})I_2(\gamt \bett \frac{\pt}{T}) + f_1 V_1 v_{2t}(\pt)(1+ \frac{\mt}{c_1 e_1 n})^{-c_1 n}
\right.\nonumber\\
& + \left.f_2 V_2 v_{2p}(\pt)(c_2^2 +\frac{\pt^2}{e_2^2})^{-\frac{n}{2}}\right].
\end{align}

$I_0$ and $I_2$ in Eqs.~(\ref{eq:F}) and (\ref{eq:v2}) are modified Bessel functions of the first kind. So, one finds Eqs.~(\ref{eq:d2N}), (\ref{eq:F}), and (\ref{eq:v2}) for hadrons \pt-spectrum and elliptic flow, respectively. They can be used to describe the corresponding data for different identified particles. There are also similar data sets for the unidentified charged particles which are a mixture of the $\pipm, \kapm, \p, \pbar$, and charged hyperons. These data have usually much larger \pt range than the ones for the identified particles. Charged-particle \pt-spectra can be defined, assuming the same spectrum for a particle and its antiparticle, by the equations~\cite{G21}
\be  \label{eq:chp}
\frac{\td^3 N_{\rm ch}}{\td\pt\td\eta\td\varphi}=2(\sum\limits_{i=\pi,\mathrm{K,p}}^{} \frac{\pt}{{\mt}_{,i}}F_{\varphi,i} + f_{hyp}\frac{\pt}{{\mt}_{,\mathrm p}}F_{\varphi,\mathrm{p}})\, ,
\ee
\be  \label{eq:chp1}
\frac{\td^2 N_{\rm ch}}{\td\pt\td\eta} = 2(\sum\limits_{i=\pi,\mathrm{K,p}}^{} \frac{\pt}{{\mt}_{,i}}F_{i} + f_{hyp}\frac{\pt}{{\mt}_{,\mathrm p}}F_{\mathrm{p}})\, .
\ee
Here, the functions $F_{\varphi,i}$ and $F_{i}$ are defined by Eqs.~(\ref{eq:Fv}) and (\ref{eq:F}), respectively.
The factors 2 account for the positive and negative particles and factors $\pt/\mt$ account for the change from rapidity to pseudo-rapidity at midrapidity ($d\eta/dy \approx \pt/\mt$). The last terms in Eqs.~(\ref{eq:chp}) and (\ref{eq:chp1}) describe approximately the small contribution of hyperons via the proton contribution, scaled by parameter $f_{hyp}$. Using Eqs.~(\ref{eq:chp}) and (\ref{eq:v2N3}) and performing integration by $\varphi$ one can find easily the analog of Eq.~(\ref{eq:v2}) for the unidentified charged particles. Note that Eqs.~(\ref{eq:d2N}), (\ref{eq:F}) and (\ref{eq:chp1}) for the \pt-spectra are similar to the corresponding ones in Ref.~\cite{G21} if $f_N=f_S=1$ and $\rho_2=0$.

It was checked that the presented generalized BWTPM is able to fit the LHC data on midrapidity hadron \pt-spectra and \vv in \pbpb and \xexe collisions with a good accuracy, similar to that achieved in Ref.~\cite{G21}. However, as in Ref.~\cite{G21}, the number of the model free parameters is too big, which reduces the model's predictive power. It is mostly due to the ten centrality-dependent parameters $V, V_1, V_2, T, k, e_2, n, f_S, r_{xy},$ and $\rho_0$ which generally can depend also on \pt, the collision system and \enn ($f_S$ depends on particle species too). 
A typical number of ALICE centrality classes for \pbpb data at $\enn=$ 2.76~TeV~\cite{alic4} and 5.02~TeV~\cite{alic5} as well as for \xexe data at $\enn=$ 5.44~TeV ~\cite{alic6} is about 10 in each case. So, to describe these combined data, one needs to fit $10(10+10+10)=300$ independent values for the mentioned ten parameters. Taking into account also the RHIC data included in the present study, this number of independent values becomes more than 600. 
To reduce the number of free parameters strongly, these ten (and other) parameters are parametrized using the charged-particle multiplicities \mch which depend on the collision system, centrality and \enn, and are well known experimentally for various AA collisions ~\cite{alic4,alic5,alic6,alic536,phob,phen1}. The key assumption of the present paper is that all the centrality dependence in BWTPM can be expressed via \mch only. It is motivated by the experimental 
observation that any hadron species has similar \pt-spectra in different AA collisions with the same \mch and \enn (see an example of such a similarity for \kstar~meson \pt-spectra in Ref.~\cite{alic2308}). To have a more general and predictive approach, I will use not the experimental values of \mch (which can be missing for some centralities), but their accurate parametrization, valid for any AA collision centrality and energy. This parametrization is presented in the next section.

\section{\label{sec3}parametrization of \mch}

\begin{figure}[ht]
\includegraphics[width=1.1\columnwidth]{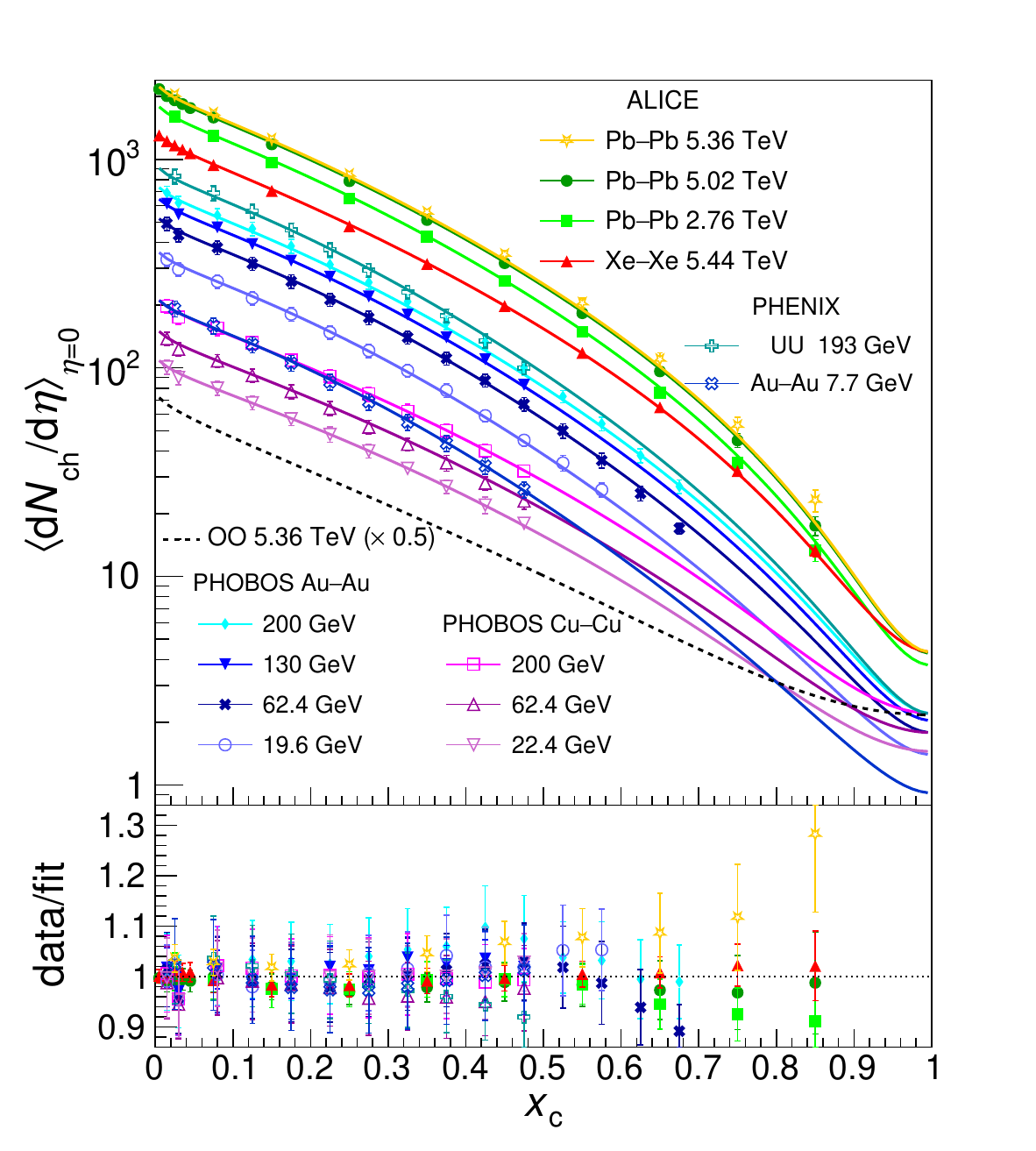}
\vskip -2mm
\caption{Fit of the midrapidity charged-particle multiplicity density data measured for various AA collisions at different centralities $x_c$ and energies \enn by the ALICE~\cite{alic4,alic5,alic6,alic536}, PHOBOS~\cite{phob} and PHENIX~\cite{phen1} experiments. The dashed line shows the prediction (multiplied by 0.5 for better visibility) for the OO collisions at $\enn=$5.36~TeV. The data/fit ratio demonstrates the quality of the fits.} \label{nchaa}
\end{figure}

\begin{figure}[ht]
\includegraphics[width=1.1\columnwidth]{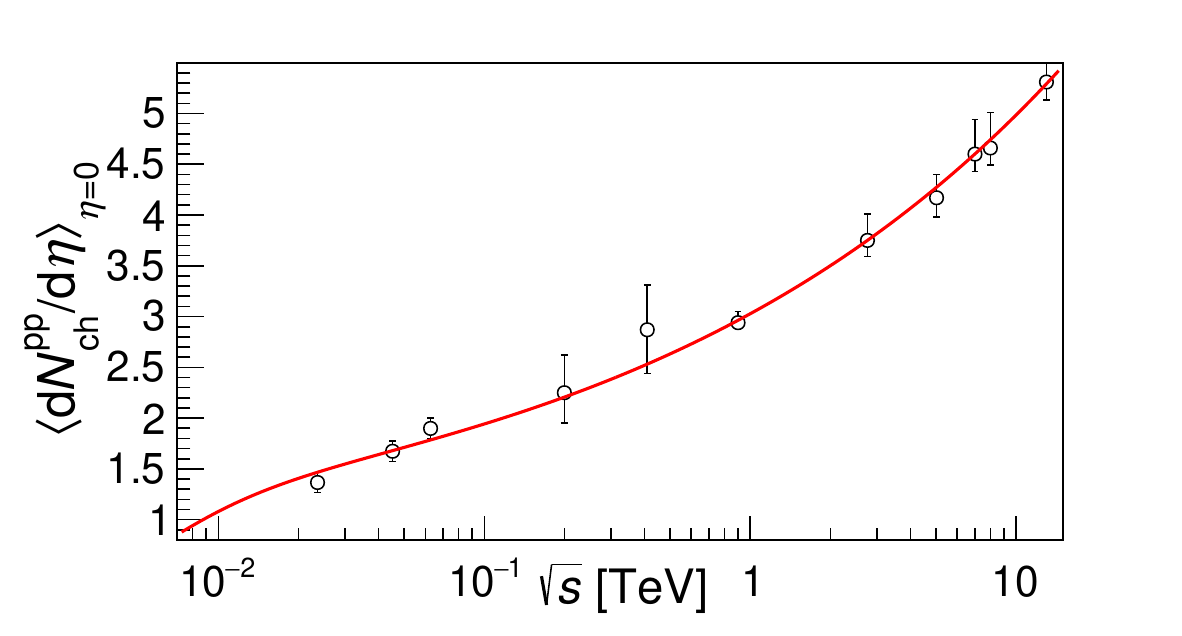}
\vskip -2mm
\caption{Fit of the midrapidity charged-particle multiplicity density in inelastic pp collisions by Eq.~(\ref{eq:dNdetapp}) as a function of \en. 
Six higher energy data points are from Refs.~\cite{alic7,alic8,alic9} and the remaining data points are from Refs.~\cite{phob,T77}.} \label{nchpp}
\end{figure}

The mean value of charged-particle multiplicity density at midrapidity \mch (defined in Eq.~(\ref{eq:mch0})) is one of the main characteristics of nucleus-nucleus collisions. Its dependence on the collision system, centrality and \enn was measured for different AA collisions at LHC~\cite{alic4,alic5,alic6,alic536} and RHIC~\cite{phob,phen1} energies. Centrality is commonly expressed in percentiles or fractions $x_c$ of the total nucleus-nucleus cross section~\cite{alic1} (centrality in \% is equal to $100 x_c$). Here I propose a simple parametrization for the \mch data~\cite{alic4,alic5,alic6,alic536,phob,phen1} versus atomic mass number A, collision centrality fraction $x_c$ and energy \enn. It relates \mch with the similar quantity $\mchpp \equiv \mdNdepp$ of the inelastic pp collisions at energy $\en=\enn$. For \mchpp I used the data measured at the LHC~\cite{alic7,alic8,alic9} as well as at lower energies~\cite{phob,T77} and parametrized their energy dependence. The \mch and \mchpp data were obtained by averaging the respective multiplicity distribution on pseudo-rapidity in the range $|\eta|<0.5$~\cite{alic4,alic5,alic6,alic536,phen1,alic7,alic8,alic9} or $|\eta|<1$~\cite{phob,T77}. 
Within the data uncertainties both ranges give similar results~\cite{alic9}. Parameters of the parametrization are obtained via a simultaneous fit of the \mch and \mchpp data using the combined fit technique of the ROOT framework~\cite{Root}, which gives $\chi^2/NDF=54.1/155$. The fit results are shown in Figures~\ref{nchaa} and~\ref{nchpp} and the parametrization has the form
\be  \label{eq:dNdeta}
\frac{\mch}{\mchpp}=1+(\textrm{A}-1)^{1+0.00013y}\frac{(1-x_c^2)^{2+0.0018y\alpha}}{(x_c^\delta+\beta)^{2.1}\gamma},
\ee

\vspace{-10pt}

\be  \label{eq:dNdetapp}
\mchpp = 0.615\frac{1+(\enn/e_0)^{0.258}}{1+(e_0/\enn)^2},
\ee
where $\alpha=\log(\frac{17.4}{\enn}), \beta=(\frac{0.0027}{\enn})^{0.054}, \delta=\textrm{A}^{0.063}, \gamma=1+1.22x_c^{0.47}, y=\textrm{A}(1-x_c)$, and $e_0 = 0.005$~TeV (everywhere in the present paper, if it is not specified otherwise, $\enn$ is defined in units of TeV). The bottom panel of  Fig.~\ref{nchaa} demonstrates that Eq.~(\ref{eq:dNdeta}) describes most of the data points with better than 10\% accuracy. However, it underestimates the ALICE recent data for the most peripheral \pbpb collisions at 5.36~TeV~\cite{alic536}. This could be due to the new centrality calibration~\cite{alic536}, since all other ALICE data are described well. One can expect that  Eq.~(\ref{eq:dNdeta}) would give accurate values of \mch also for new AA collisions. The result for oxygen--oxygen (OO) collisions, carried out at the LHC in July 2025, is also shown in Fig.~\ref{nchaa}. Note that for $\textrm{A}=1$ or for the most peripheral collisions, when $x_c \to 1$, the right-hand side of Eq.~(\ref{eq:dNdeta}) equals unity. For most central collisions, when $x_c \to 0$, one obtains a very simple formula for the maximal values of \mch versus A and \enn:
\be  \label{eq:dNdetamax}
\frac{\mch^{max}}{\mchpp}=1+(\textrm{A}-1)^{1+0.00013\textrm{A}}\bigl(\frac{\enn}{0.0027}\bigr)^{0.1134}.
\ee
For example, $\mch^{max}=$~2404 (2457) for \pbpb collisions at $\enn=$5.02 (5.36)~TeV.
It is worth noting that Eq.~(\ref{eq:dNdeta}) should be used with caution since it gives the same \mch for collisions of isobars of the same A~\cite{starIs1}.

To my knowledge, there is no generic parametrization in the literature similar to Eq.~(\ref{eq:dNdeta}). The existing parametrizations of \mch versus $x_c$ (or versus the  number of participant nucleons \Npart) are valid only for a particular AA collision system~\cite{phob,star1}. Of course, there are many sophisticated models describing the \mch data, most of which use the hydrodynamical approach for particle production in AA collisions (see, e.g., Ref.~\cite{Gia1}). However, such models are much more complex with respect to the simple parametrization of Eq.~(\ref{eq:dNdeta}) and their accuracy is worse. 

\vspace{-2em}

\section{\label{sec4}parametrization of BWTPM}

Here I present parametrizations for the model parameters versus \mch (given by Eq.~(\ref{eq:dNdeta})), A, \enn, and the particle mass and \pt. The parametrizations were chosen to be as simple as possible by examining the behavior of the model parameters versus the collision system, centrality, and energy, and using the results of Ref.~\cite{G21} and other studies. For normalization parameters $V, V_1$, and $V_2$, to describe their strong increase with centrality~\cite{G21} and taking into account the experimental fact that the hadron \pt-integrated yields increase linearly with \mch at $\mch>100$~\cite{Cley1}, the following parametrizations are assumed:
\be  \label{eq:Vs}
V=a_1\mch/(1+f_r\mch^{1/3}\frac{m}{\mt})^{a_4}, ~V_1=a_2\mch,~V_2=a_3\mch.
\ee
Here and throughout the present paper the parameters $a_i, i=1,2,3,...$ are fitting constants. Parameter $f_r$ is nonzero for \kstar and \rhz only, and accounts for the suppression of the yields of these short-lived resonances at low \pt due to the rescatterings of their decay products in the hadronic medium. The suppression is stronger for more central collisions, and scales with $\mch^{1/3}$~\cite{alic1}.

The decrease of parameters $T$ and $k$ with increase of centrality and of parameters $e_2$ and $n$ with increase of centrality and \enn~\cite{G21} can be parametrized as follows:

\be  \label{eq:Tk}
T=a_5/(1+a_6\log\mch), ~k=a_{7}/(1+a_{8}\mch),
\ee


\be  \label{eq:e2}
e_2=a_{9}/(1+a_{10}\log\enn)-a_{11}\log\mch,
\ee

\vspace{-15pt}

\be  \label{eq:n}
n=h_n[\frac{a_{12}}{1+a_{13}y\log\enn} + \frac{1+(a_{14}/\enn)^{2y}}{(\mch/a_{15})^{a_{16}/(1+x^2)}}],
\ee
where $x=\pt/(a_{17}\log(10^3\enn)), y=(\pt/\mt)^{a_{18}(c_2-1)}$ , $h_n$ differs from unity only for the strange and charmed baryons and \jpsi (\enn is in units of TeV). Parameter $c_2$ is given below in Eq.~(\ref{eq:c2}). The term with parameter $a_{14}$ is introduced to describe the RHIC data. 
It is significant at LHC energies for $\pt \ll m$ only. Note that $n$ becomes independent of \mch at high \pt where $x > 1$.

The factor $f_S$ is equal unity for ordinary hadrons and increases with decreasing centrality (or \mch) for other hadrons. It is greater for hadrons containing larger number of strange and charm (anti)quarks, which are denoted by $n_s$ and $n_c$, respectively. For example, $n_s=1$ for K and \etz~ mesons, and $n_s=2$ for the $\phi$ meson. A simple parametrization for $f_S$ in AA collisions is found
\be  \label{eq:fs}
f_S=1+\frac{[a_{19}n_s+a_{20}n_c+a_{21}|B|(n_s+n_c)]^2}{(1+\pt^2/a_{22}^2)\sqrt{\mch+a_{23}}},
\ee
which describes well the enhancement features of all considered hadrons containing at least one strange or charm (anti)quark ($n_s =n_c = 0$ for ordinary hadrons). The term proportional to the hadron baryon number $B$ is introduced to better describe the production of hyperons. The same parameter $a_{21}$ for strange and charmed baryons is assumed since the data used for the latter case are very limited.
For the hadrons containing bottom quarks, not considered in the present study, Eq.~(\ref{eq:fs}) can be modified accordingly. Note that $f_S$ tends to unity at $\pt>a_{22}$. Hence, the ratios of different particle yields in the present model are independent of centrality at high \pt, as in the data~\cite{alic1}.

It is known that the \pt-averaged \vv increases when going from central to peripheral collisions, and then decreases for very peripheral collisions~\cite{alic1,alic3}. Also, the eccentricity \ecc of the transverse geometry of the hadronic medium behaves similarly versus collision centrality. 
When going from central to peripheral collisions, this geometry goes from circular to elliptical and \ecc increases. But the geometry goes again to circular and \ecc decreases at very peripheral collisions, where on average only one nucleon-nucleon collision accurs~\cite{HL99,Heinz1}. So one can assume $\vv \sim \ecc$.
This parametrization of $r_{xy}$  ensures such a centrality dependence for \ecc (given by Eq.~(\ref{eq:ecc2}))
\be  \label{eq:rxy}
r_{xy}=1-\frac{a_{24}\textrm{A}^{a_{25}}}{1+a_{26}/x+a_{27}x^{a_{29}}+a_{28}[(1-f_d)x]^{a_{30}}},
\ee
where $x=\textrm{A}^{a_{25}}(z_c+y)/(1+y)$, A is the atomic mass number, $z_c=\mch/\mch^{0\textrm{-}5\%}$ is the normalized multiplicity over the multiplicity of the 0-5\% most central collisions $\mch^{0\textrm{-}5\%}$ (determined by Eq.~(\ref{eq:dNdeta}) at $x_c=0.025$) and its correction $y=(a_{31}/\enn)^{a_{32}}z_c$ is needed mainly for the fit of \vv data at RHIC energies. 
Parameter $f_d$ is introduced to describe a slightly different centrality dependence of \vv data in \xexe collisions  compared to all other collision systems considered in present study for which $f_d=0$ is used. The difference  is related to the non-spherical (deformed) shape of the $^{129}$Xe nucleus~\cite{alic10}.

To describe the centrality and energy dependence of the hadrons \pt-spectra and also of the charged-particle average transverse momentum \mpt~\cite{alic2211,cms2401}, particularly its local minimum at $z_c\approx1.05$ and rise at $z_c>1.05$ for ultracentral collisions~\cite{cms2401}, I parametrized $\rho_0$ as
\begin{align} \label{eq:rho0}
& \rho_0=\frac{1}{f_{uc}}\bigl[\frac{a_{33}\enn^{a_{34}}}{1+f_d\frac{\mch}{a_{35}}}-a_{36}\log^2(\frac{z_c}{a_{37}})\bigr], \nonumber\\
& f_{uc}=\biggl[1-\frac{a_{38}}{z_c}\sqrt{\frac{2}{\pi}}\frac{\exp(-\frac{(z_c-a_{39})^2}{2a_{38}^2})}{\textrm{erfc}(\frac{z_c-a_{39}}{\sqrt{2}a_{38}})}\biggr]^{a_{40}}.
\end{align}
\vskip-0.5em
\noindent
Factor $f_{uc}$~\cite{cms2401,Gia2} is close to unity at $z_c\! \le \!a_{38}\! \approx\!1.1$ and decreases at $z_c\! > \!a_{39}$. So, it increases $\rho_0$ at $z_c\! > \!a_{39}$ and therefore increases \mpt for ultracentral events as in Refs.~\cite{cms2401,Gia2}. Parameter $a_{35}$ is used to describe the observed slightly different centrality dependence of \mpt in \xexe and \pbpb collisions, which is due to the non-sphericity of $^{129}$Xe nuclei~\cite{alic2211}.

So, the parametrizations in Eqs.~(\ref{eq:Vs})--(\ref{eq:rho0}) allow us to determine the aforementioned (in Sec.~\ref{sec2}) more than 600 independent values for $V, V_1, V_2, T, k, e_2, n, f_S, r_{xy},$ and $\rho_0$ by means of about 40 constants. 

I also parametrized other ingredients of BWTPM. The energy dependences of parameter $f_{hyp}$ from Eqs.~(\ref{eq:chp}) and~(\ref{eq:chp1}) and of the global normalization $f_N$ are given as
\be \label{eq:fN}
f_N = a_N[1+(a_{41}/\enn)^{a_{42}}], ~f_{hyp}=\enn/a_{43},
\ee
\vskip-0.5em
\noindent
where $a_N=1$ is assumed for \pbpb collisions at $\enn=$ 5.02~TeV, and for other used data sets the values of $a_N$ are defined relative to this one.
As mentioned in Sec.~\ref{sec2}, it is assumed that the parameter $\rho_2$ as well as the functions $v_{2t}(\pt)$ and $v_{2p}(\pt)$ are proportional to \ecc
in order to ensure the relation $\vv \sim \ecc$~\cite{Vol3,Oll2,Heinz1,alic3}. A weak energy dependence of $\rho_2$ is parametrized as
\be  \label{eq:rho2}
\rho_2=\ecc a_{44}/\enn^{a_{45}}.
\ee
\vskip-0.5em
\noindent

Functions $v_{2t}(\pt)$ and $v_{2p}(\pt)$ are responsible mainly for the description of the high-\pt behavior of elliptic flow \vv. For their dependence on \pt, \enn, and \mch I use
\be  \label{eq:v2tp}
v_{2t}(\pt)=\ecc\frac{a_{46}}{y}\frac{x^z}{1+t^2}, ~~v_{2p}(\pt)=\ecc\frac{a_{47}}{y}\frac{x^z}{1+t^2},
\ee
\vskip-0.5em
\noindent
where $t=p_2/\pt$, $x=$~GeV/\mt, $y=1+a_{48} m/\enn$, $z=a_{49} - a_{50}\log{z_c}$, 
and parameter $p_2$ depends on the particle type. The latter can be expressed via the particle mass $m$ as follows
($m_\pi$ is the $\pi^{\pm}$ mass)
\be  \label{eq:p2}
\hskip-3mm\frac{p_2}{\textrm{GeV}/c} \!=\! \left\{
\begin{array}{ll}
a_{51}+a_{52}(1-\frac{m_\pi}{m})\log(\frac{a_{53}}{m})\; \textrm{for mesons},\\
a_{54}+a_{55}\exp(\frac{a_{56}}{m}) \qquad\qquad  \textrm{for baryons},\!\!\\
a_{57} \qquad\qquad\qquad\qquad\qquad\;\; \textrm{for } \jpsi.
\end{array}
\right. 
\ee
\vskip-0.5em
\noindent
The parametrization for mesons is inaccurate for \jpsi.

The particle type dependence of parameter $c_2$ can be expressed by a simple relation
\be  \label{eq:c2}
c_2 = 1+\frac{m}{\textrm{GeV}/c^2}\left\{
\begin{array}{ll}
a_{58}H(m-m_\pi) \quad  \textrm{for mesons},\\
a_{59}H(m-m_\p) \quad  \textrm{for baryons},
\end{array}
\right. 
\ee
\vskip-0.5em
\noindent
where $m_\p$ is the proton mass and the step function $H$ is defined as $H(x)=0(1)$  for $x\leq(>)~0$. So $c_2=1$ for pions and (anti)protons. Larger $a_{59}$ with respect to $a_{58}$ is nedded to describe the baryon yields. Note that the different parametrizations for mesons and baryons in Eqs. ~(\ref{eq:p2}) and~(\ref{eq:c2}) indicate the difference in production mechanisms of these hadrons containing two and three (anti)quarks, respectively.

Parameter $e_1$ shows a centrality dependence significant mainly for pions. It can be described by
\vspace{-5pt} 
 \be  \label{eq:e1}
e_1=\frac{\tilde{e}_1}{1+\sqrt{f_\p/\enn}}(1+\frac{m_\pi}{m} \frac{a_{60}}{\mch+a_{23}}),
\ee
\vskip-0.5em
\noindent
where parameter $\tilde{e}_1$ depends on the particle species. Parameter $f_\p$ is nonzero for the (anti)proton only to better describe its \pt-spectra at RHIC energies. Additional normalization parameter $f_{\rm D}$ is used for \Dpm mesons to describe their lower yield with respect to the \Dzero~meson.

\section{\label{sec5}Fitted data and parameters}
\vspace{-0.5em}

Thus, all the ingredients of the BWTPM are presented in previous sections. To determine the values of its free parameters, a global simultaneous fit is performed of the combined data on various hadrons midrapidity \pt-spectra, \mpt, and \vv for different centrality classes in \pbpb collisions at energies $\enn=2.76$ and 5.02~TeV~\cite{alic5,alic2211,cms2401,alic11,alic12,atl1,cms1,cms2,alic13,alic14,alic15,alic16,alic17,cms3,atl2,alic18, alic19,alic20,alic21,cms4,cms5,alic22,alic23,alic24,alic25,alic26,alic27,alic28,alic1903} and in \xexe collisions at 5.44~TeV~\cite{alic10,alic29,alic30,alic31}. To check the validity of BWTPM for other colliding nuclei and at much lower energies, this fit also includes the \pt-spectra and \vv for \pipm, \kapm, ~and \p(\pbar) measured in \mbox{Au--Au} collisions at $\enn= 200$~GeV~\cite{phen2,star2}, and the \pt-spectra for \pipm and \kapm ~measured in Au--Au collisions at 39 and 27~GeV~\cite{star3}, in UU collisions at 193~GeV~\cite{star4}, and in Cu--Cu collisions at 
200~GeV~\cite{brah}. In the last four cases the proton data not included since they are not corrected for the contribution from weak decays of hyperons. 
The resulting ratio $\chi^2/NDF$ for this global fit (in the ROOT framework~\cite{Root}) is 6738/13408. Values for the fitting parameters are given in Tables I--III. 
Parameters $f_{BW}, f_1$, and $f_2$ for \pipm and \pizero are fixed equal to unity. Note that the fit uses the data of prompt ${\rm D}$ and \jpsi mesons and $\Lambda_c$ baryons, not including contributions from decays of heavier hadrons, containing $b$-quarks. To describe the \pt-spectra in \xexe collisions~\cite{alic29,alic30}, the right-hand side of Eq.~(\ref{eq:chp1}) is scaled by a factor of 1.07 to account for a small normalization difference between the measured \pt-spectra of identified~\cite{alic29} and unidentified charged particles~\cite{alic30}.
Note also that parameters $a_{24}$--$a_{32}$ and $a_{44}$--$a_{57}$ are resposible mainly for the elliptic flow. Only some of them entering in Eq.~(\ref{eq:rho2}) influence the azimuthally integrated \pt-spectra due to the $\rho_2$ contribution into the radial flow velocity Eq.~(\ref{eq:bt}). This contribution is small and decreases with increasing energy, according to Eq.~(\ref{eq:rho2}). So, its influence on \pt-spectra is negligible at high energies and becomes sizable (at most about several \% at high \pt) only for Au--Au collisions at low energies $\enn=27$ and 39~GeV. 
It should be noted that all the parameters listed

\begin{widetext}

\begin{table*}[ht]
   \caption{Particle-type dependent model parameters.} \label{tab1}
  \begin{tabular}{*{14}{c}}
    \colrule
   parameter \;& \pipmns, \pizero & \kapm, \kzero & \p, \pbar & \etz & \rhz &  \kstar, \akstar & \; \phm &\lmb, \almb  & \X, \Ix &\; \Om, \Mo &\; \Dzero, \Dpm & \jpsi & \lmbc,\almbc \\
    \colrule
$f_{BW}$ & 1 & 0.993 & 3.76 & 1.26 & 2.18 & 2.23 & 1.66 & 10.96  & 7.08 & 8.29 & 192 & 1500 & 3600\\
$f_1$       & 1 & 0.085  &  0.00015 &  0 & 0 & 0 & 0 & 0 & 0 & 0  &  0
 & 0 & 0 \\
$f_2$ & 1 & 0.528  & 0.041 & 0.490 & 0.360 & 0.163 & 0.117 & 4.70  & 1.64  & 0.428 & 0.735 & 0.111 & 6.0 \\
$c_1$        &  1.30 &  1.43 & 3.06 & - &  - &  - &  - &  - & -   & -   & -  & - & -  \\
$\tilde{e}_1$ [GeV] & 0.065  & 0.099  & 0.361 & - & - &  - &  - & - &  - & -  & - & - & -  \\
$h_n$        &  1 &  1 & 1 & 1 & 1 & 1 & 1 & 1.30 & 1.30   & 1.30 & 1 & 1.14 & 1.20 \\
    \colrule
   \end{tabular}
\end{table*}

\vspace{-1.0em}

\begin{table*}[ht]
   \caption{Parameters $a_1$--$a_{60}$ of Eqs.~(\ref{eq:Vs})--(\ref{eq:e1}), $f_{r}$ (is nonzero for \rhz and \kstar only) of Eq.~(\ref{eq:Vs}) , $f_{d}$ (is nonzero for Xe--Xe only) of Eqs.~(\ref{eq:rxy}) and ~(\ref{eq:rho0}),  $f_{\p}$ (is nonzero for \p~and \pbar~only) of Eq.~(\ref{eq:e1}), and $f_{\rm D}$ for normalization of \Dpm mesons.} \label{tab2}
  \begin{tabular}{*{13}{c}}
    \colrule
$a_1[{\rm fm}^3]$ & $a_2[{\rm fm}^3]$ & $a_3[{\rm fm}^3]$ & $a_4$ & $a_5$  [GeV]& $a_6$  & $a_7$ & $a_8$ & $a_9$ [GeV] & $a_{10}$ & $a_{11}$  [GeV] & $a_{12}$ & $a_{13}$ \\
6.15 & 17.62 & 0.015  & 2.0 & 0.1354 & 0.020 & 2.61 & 0.012 & 1.477 & 0.015  &  0.059 & 7.66 & 0.080 \\
    \colrule
$a_{14}$ [TeV] & $a_{15}$ & $a_{16}$ & $a_{17}$ [GeV] & $a_{18}$ & $a_{19}$ & $a_{20}$ & $a_{21}$ & $a_{22}$  [GeV] & $a_{23}$ & $a_{24}$ & $a_{25}$  & $a_{26}$  \\
0.027 & 10.85 & 0.80 & 35.0 & 3.60  & 0.92 &  2.50 &  0.53 & 24.0 & 5.0 & 0.341 & 0.070 & 0.0042 \\
    \colrule
$a_{27}$ & $a_{28}$ & $a_{29}$ & $a_{30}$ & $a_{31}$ [TeV] & $a_{32}$ & $a_{33}$ & $a_{34}$ & $a_{35}$ & $a_{36}$ & $a_{37}$ & $a_{38}$ & $a_{39}$ \\
0.782 & 0.159 & 1.56 & 6.63  & 0.166  & 1.80  & 0.984  & 0.083 & 5100 & 0.014 & 0.81 & 0.014& 1.103 \\
    \colrule
$a_{40}$ & $a_{41}$ [TeV] & $a_{42}$ & $a_{43}$ [TeV] & $a_{44}$ & $a_{45}$ & $a_{46}$ & $a_{47}$ & $a_{48}$ & $a_{49}$ & $a_{50}$ & $a_{51}$ & $a_{52}$ \\
0.301 & 0.02 &  0.52 & 21.5 & 0.386 & 0.131 & 6.13 & 2.13 &  0.166 &  0.689 & 0.042 & 3.04 & 1.14 \\
    \colrule
$a_{53}$ [GeV] & $a_{54}$ & $a_{55}$ & $a_{56}$ [GeV] & $a_{57}$ & $a_{58}$ & $a_{59}$ & $a_{60}$ & $f_{r}$, \rhz & $f_{r}$, \kstar & $f_{d}$, Xe--Xe & $f_{\p}$ [TeV] & $f_{\rm D}$ \\
0.76 & 1.60 & 0.05  & 3.63 & 2.67 & 0.305 & 0.595 & 1.70 &  0.112 &  0.112 & 0.133 & 0.009 & 0.45 \\
    \colrule
   \end{tabular}
\end{table*}

\end{widetext}

\begin{table}[ht]
\centering
   \caption{Normalization parameter $a_N$ defined in Eq.~(\ref{eq:fN}) for all considered collision systems. Corresponding \enn values are given in units of TeV.} \label{tab3}
  \begin{tabular}{*{9}{c c c|c|c c c|c|c}}
    \colrule
 & \multicolumn{2}{c|}{Pb--Pb} & \multicolumn{1}{c|}{Xe--Xe} & \multicolumn{3}{c|}{Au--Au} & \multicolumn{1}{c|}{UU} & \multicolumn{1}{c}{Cu--Cu}  \\ 
\enn & 5.02 & 2.76 & 5.44  & 0.2  & 0.039 & 0.027 & 0.193 & 0.2\\
$a_N$ & 1 & 1.044 & 0.986  & 1.008  & 1.003 & 0.973 & 1.110 & 0.965\\
    \colrule
   \end{tabular}
\end{table}

\noindent
in Tables I--III have reasonable fitting uncertainties and are important for describing the data. For instance, the small parameter $a_{26}=0.0042\pm 0.0003$ is necessary for fitting \vv for peripheral collisions. The global fit with $a_{26}=0$ has $\chi^2 = 7140$ and significantly overestimates the \vv data (shown in Fig.~\ref{v2vsMxc}) at $\mch < 40$ and $x_c > 0.7$.

\section{\label{sec6}Discussion of the results}

\subsection{\label{ssec1}Hadrons \pt-spectra, integrated yields, and \mpt}

\begin{figure}[ht]
\includegraphics[width=1.1\columnwidth]{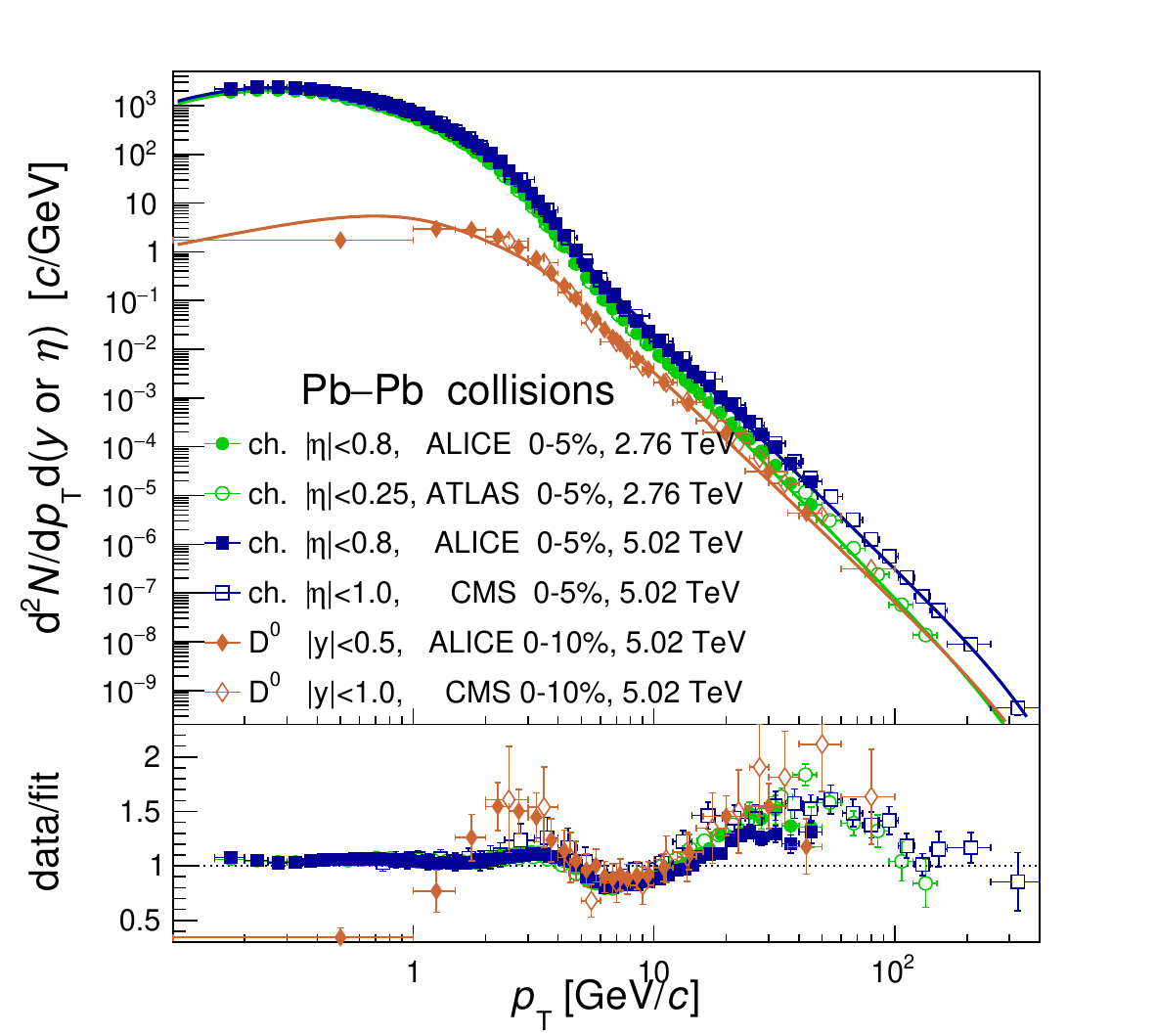}
\vskip -2mm
\caption{Fit of \pt-spectra with highest \pt reached in \pbpb collisions for charged particles at \enn$=$2.76~TeV~\cite{alic12,atl1} and 5.02~TeV~\cite{alic12,cms1} and for \Dzero\, at \enn$=$5.02~TeV~\cite{cms2,alic19}.}
\label{chD0}
\end{figure}

First, I present several plots to illustrate the results and quality of the \pt-spectra fits. Figure~\ref{chD0} displays the fits of different data sets with highest \pt reached in central \pbpb collisions. To demonstrate the quality of the fits, the data points have been divided by the corresponding fit function values, and these ratios are also plotted in the bottom panel. Generally, the quality is good within the data uncertainties. Figures 4, 5, and 6 display the main fits of the centrality dependent spectra for the most abundantly produced pions, kaons, and protons. The fit

\onecolumngrid  

\begin{figure}[H]
\includegraphics[width=1.0\columnwidth]{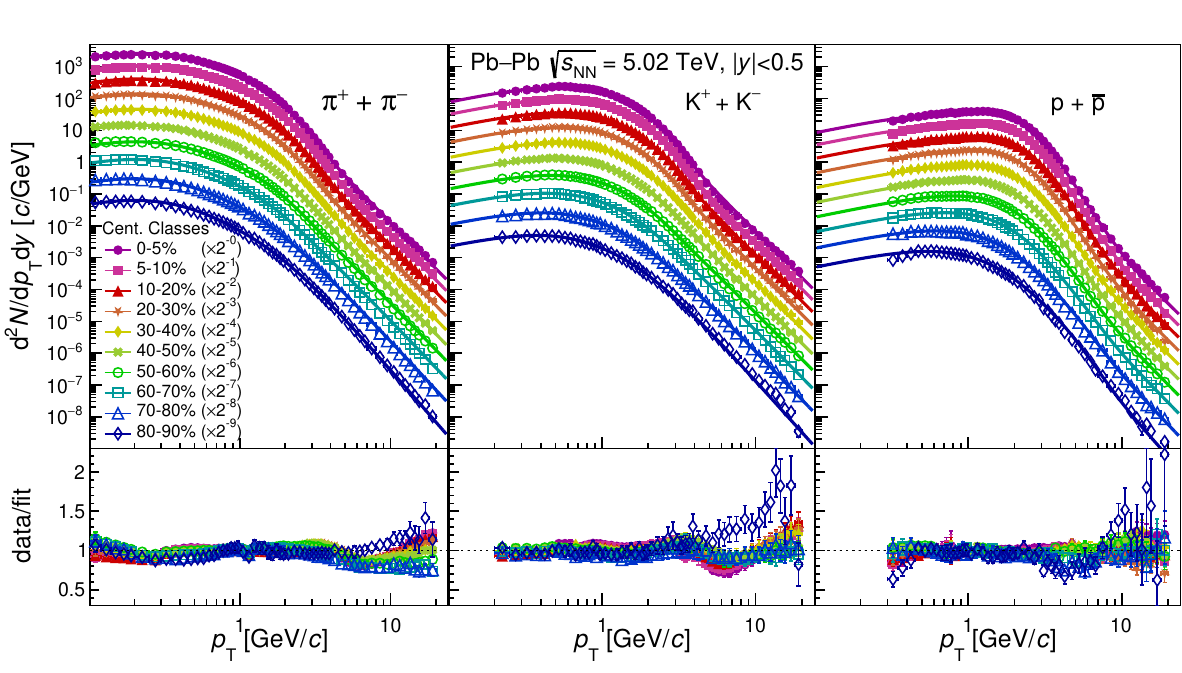}
\vskip -2mm
\caption{Fit of pion, kaon, and proton \pt-spectra at $|y| < $ 0.5 for different centrality classes (centrality intervals) in \pbpb collisions at \enn $=$ 5.02~TeV~\cite{alic5}. 
The data points and fitting curves in the top panels are scaled, for better visibility, by the numbers given in the parentheses.}
\label{idpPbPbvsM}
\end{figure}

\begin{figure}[ht]
\includegraphics[width=1.0\columnwidth]{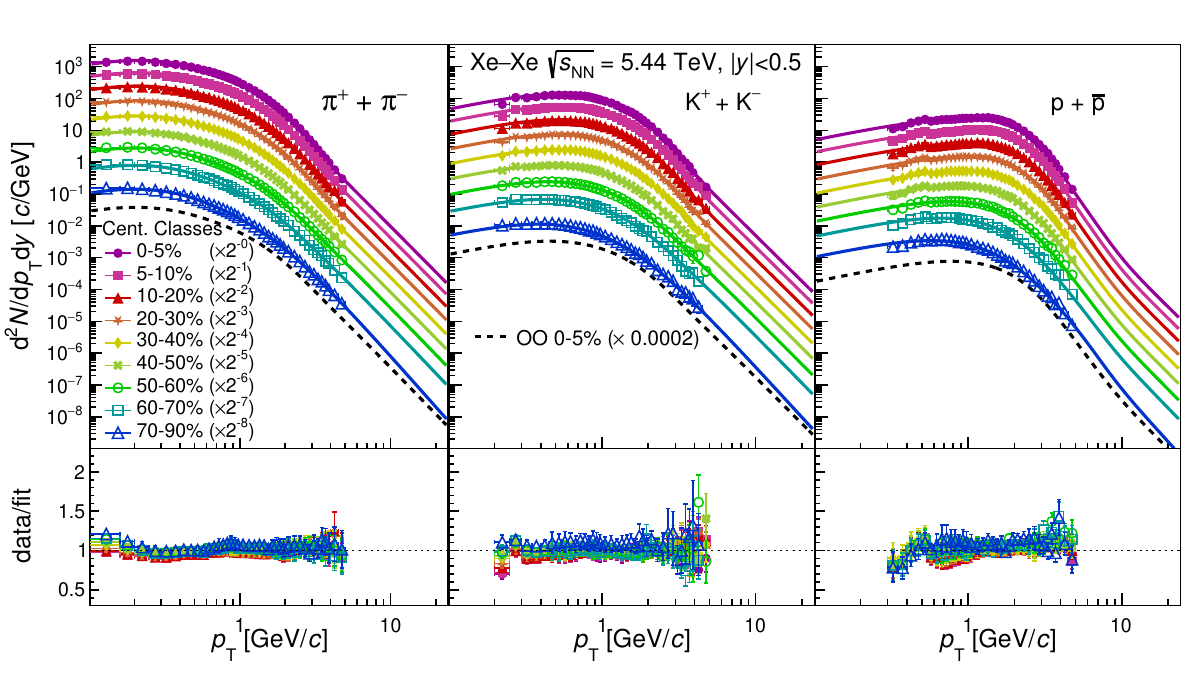}
\vskip -2mm
\caption{Similar to the Fig.~\ref{idpPbPbvsM} but for \xexe collisions 
at \enn $=$ 5.44~TeV~\cite{alic29}. In addition, the dashed lines in the top panels show the predictions (scaled down by 0.0002 for better visibility) for 0-5 \% centrality OO collisions at $\enn=$5.36~TeV.} 
\label{idpXeXevsM}
\end{figure}

\begin{figure}[H]
\includegraphics[width=1.0\columnwidth]{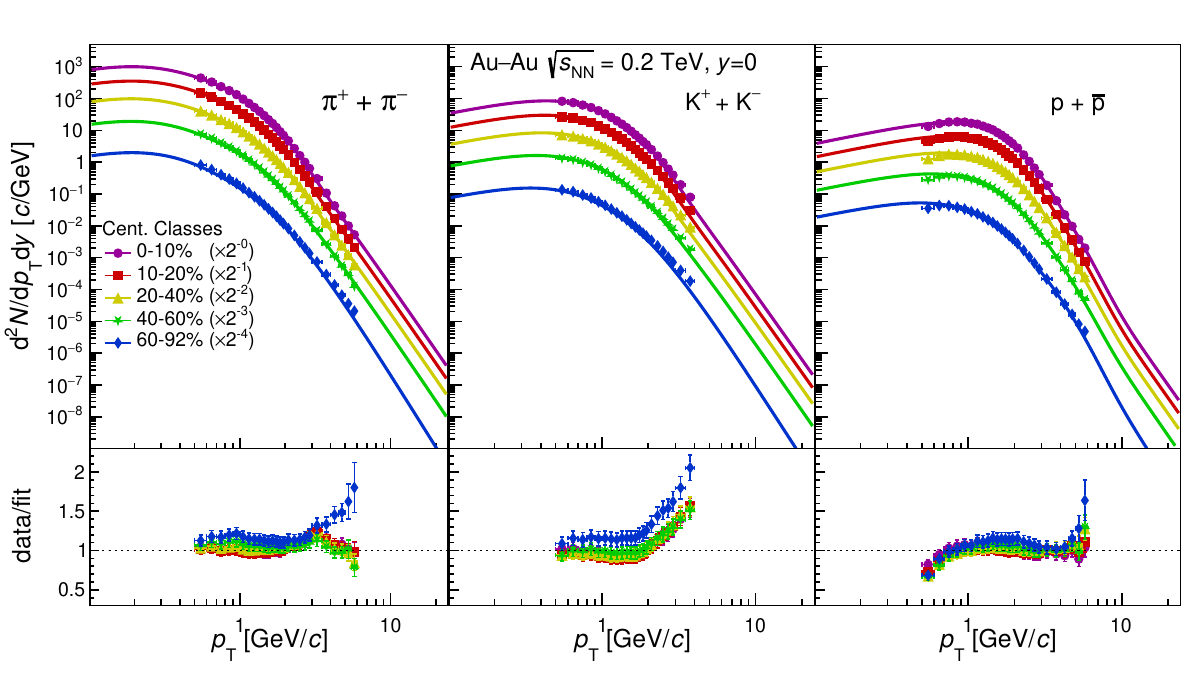}
\vskip -2mm
\caption{Similar to the Fig.~\ref{idpPbPbvsM} but for Au--Au collisions 
at \enn $=$ 0.2~TeV~\cite{phen2}.} 
\label{idpAuAuvsM}
\end{figure}

\begin{figure}[H]
\includegraphics[width=1.03\columnwidth]{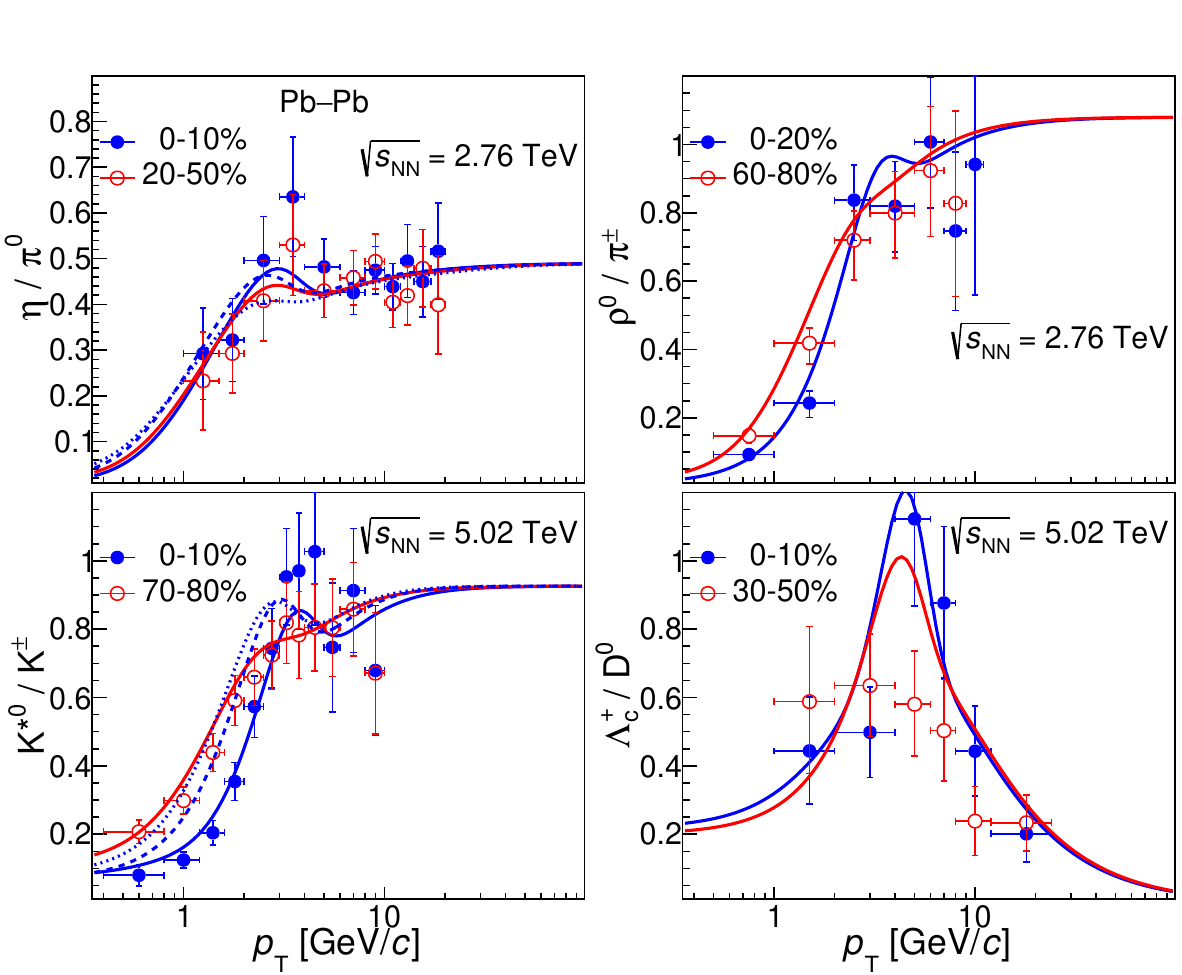}
\caption{Comparison of the model calculations for the ratios of different particle \pt-spectra with the data of central and peripheral \pbpb collisions at \enn $=$ 2.76 and 5.02~TeV~\cite{alic14,alic25,alic26,alic27}. The dashed (dotted) lines show the ratios for 0-10\% centrality Au--Au (Cu--Cu) collisions at \enn $=$ 200~GeV.} \label{RparpT}
\end{figure}

\begin{figure}[H]
\includegraphics[width=1.03\columnwidth]{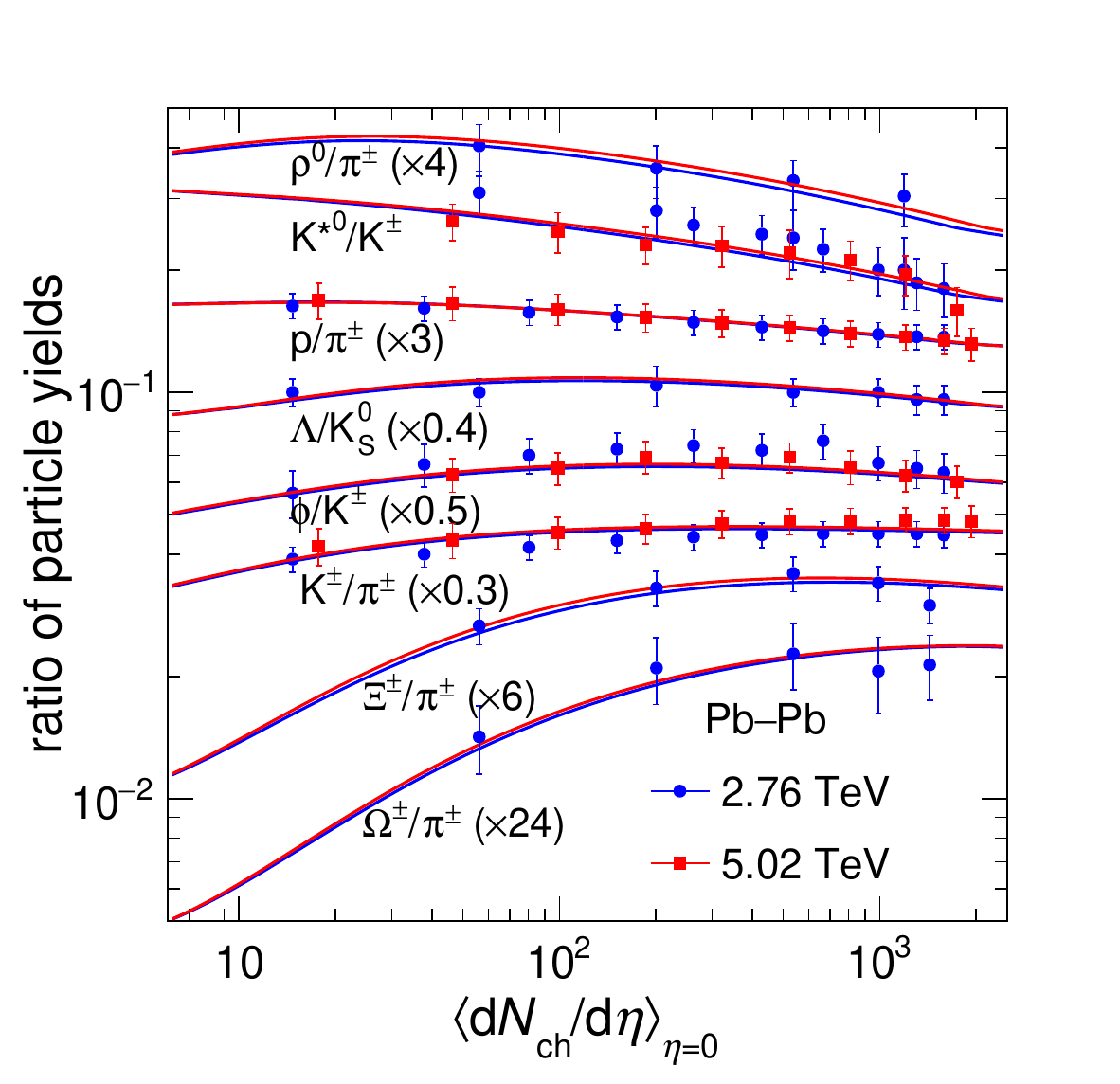}
\caption{Comparison of the model calculations for the ratios of different particle \pt-integrated yields as a function of the charged-particle multiplicity density with the measurements in \pbpb collisions at \enn $=$ 2.76 and 5.02~TeV~\cite{alic4,alic5,alic14,alic15,alic16,alic17}.} \label{RparY}
\end{figure}

\twocolumngrid  

\noindent
quality is almost always very good. This is true also for the fits of \pt-spectra of all considered other particle species. The predictions for the 0-5\% most central OO collisions (using $a_N = 1$) are shown in Fig.~\ref{idpXeXevsM}. Note that Figs.~\ref{chD0} and \ref{idpPbPbvsM} repeat the similar ones of Ref.~\cite{G21} to show that the present version of BWTPM with much fewer fitting parameters gives similar fit quality. Also, it is checked that the relative contributions of the BW, Ts, and Po components of the present model to the \pt-spectra are similar to the ones given in Ref.~\cite{G21}. As shown in Table I, the contribution of the Ts-component is nonzero only for pions, kaons, and protons, which have significant feedback from the resonance decays.\\
\indent
Then,  Fig.~\ref{RparpT} shows a good agreement of the model calculations (using the parameter values given in Sec.~\ref{sec5}) for the ratios of \pt-spectra of different hadrons with the ALICE data obtained in central and peripheral \pbpb collisions. These ratios, like the other ones given in Ref.~\cite{G21}, demonstrate more clearly the effect of hadron production mechanisms on the \pt-spectra. The peaks at \pt $\approx$ 2--7~\gevc are related to the radial flow, which is stronger for heavier particles and more central collisions. The ratios with the short-lived \kstar~and \rhz resonances show the suppression of their yields 
for low \pt and central collisions, according to Eq.~(\ref{eq:Vs}).
To demonstrate the collision system and energy dependence of particle ratios, the left panels of Fig.~\ref{RparpT} show the model calculations for 0-10\% centrality Au--Au (dashed lines) and Cu--Cu (dotted lines) collisions at \enn $=$ 200~GeV. Both dependencies are not strong and appear at moderate \pt.
Note that the hadron \pt-spectra in Eq.~(\ref{eq:F}) have a power-law behavior $\propto \pt^{2-n}$ at high \pt. Hence, the ratios for all hadrons, which have $h_n=1$ in Table I, reach a plateau (independent of the collision system, centrality, and \enn) at high \pt where parameter $n$ in Eq.~(\ref{eq:n}) depends only on \enn. The plateau is simply defined by the ratio of factors $(2J+1)f_2$ with $f_2$ given in Table I. It is true also for the ratios between hyperons that have the same $h_n$. For example, the ratio \etz/\pizero reach a plateau value of 0.49, which is compatible with the measurement $0.457\pm0.013\pm0.018$~\cite{alic25} and with the empirical fit result $0.487\pm0.024$~\cite{Ren21}.\\
\indent
Figure~\ref{RparY} displays the ratios of the \pt-integrated yields of various hadrons as a function of the charged-particle multiplicity \mch (defined by Eq.~(\ref{eq:mch0})) in \pbpb collisions. It is well known that such ratios have smooth behavior versus \mch and are almost independent of the collision system and energy~\cite{alic1,alic10}. Corresponding ratios in \xexe collisions at \enn $=$ 5.44~TeV~\cite{alic10} are not shown in Fig.~\ref{RparY} for clarity. The model curves in this case, as well as for OO collisions at \enn $=$ 5.36~TeV, coincide practically with the ones for \pbpb at \enn $=$ 5.02~TeV. The model describes well all the presented ratios. In particular, it describes the observed enhancement of yields of hadrons containing strange (anti)quarks with respect to pion yields when going from peripheral to central collisions. Enhancement is stronger for hadrons containing more strange (anti)quarks, according to Eq.~(\ref{eq:fs}). Figure~\ref{RparY} displays also the suppression of \kstar~and \rhz resonance yields when going from peripheral to central collisions. It is the manifestation of the low-\pt suppression of these resonances (see Fig.~\ref{RparpT}) due to the rescatterings of their decay products in the hadronic medium. 
The $\phi$ resonance does not have such a suppression since it lives

\begin{figure}[ht]
\includegraphics[width=1.03\columnwidth]{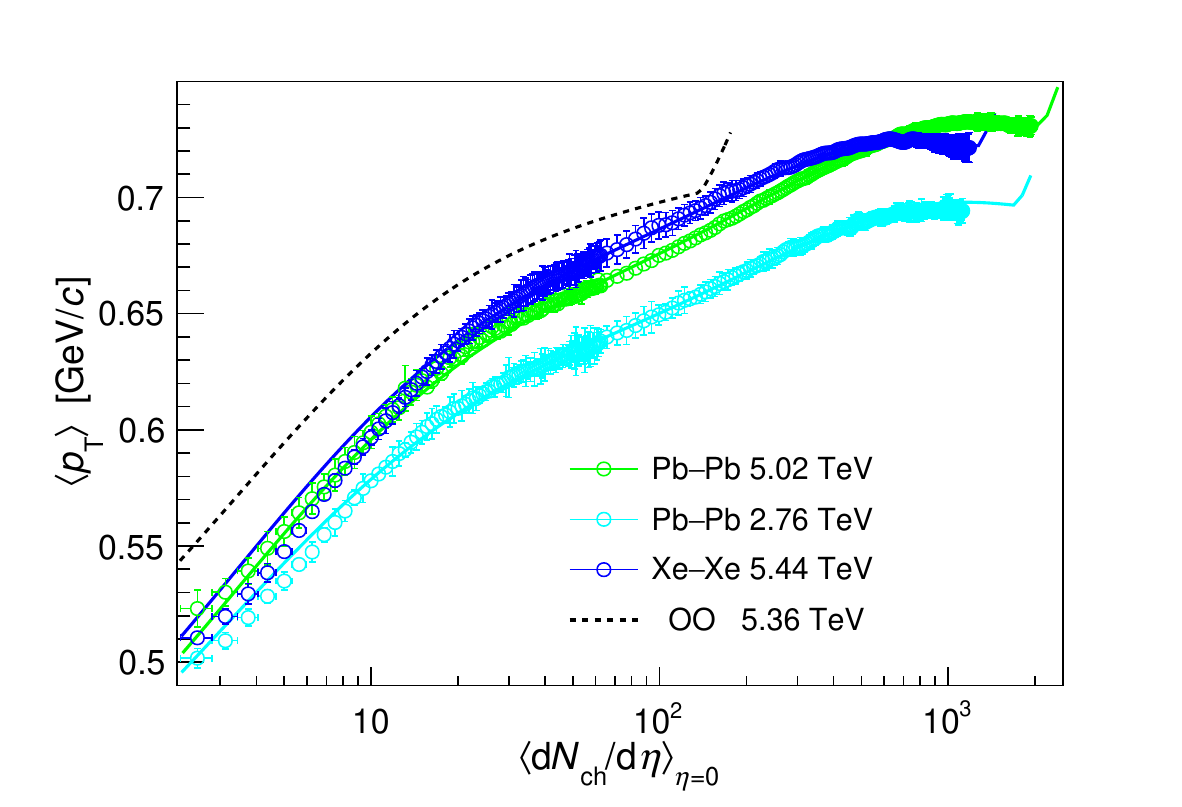}
\caption{Fit of charged particles \mpt as a function of their multiplicity density in \pbpb collisions at \enn $=$ 2.76 and 5.02~TeV, and in \xexe collisions 
at \enn $=$ 5.44~TeV~\cite{alic2211}. Dashed line shows the prediction for OO collisions at $\enn=$5.36~TeV.} \label{meanpT}
\end{figure}

\begin{figure}[ht]
\includegraphics[width=1.03\columnwidth]{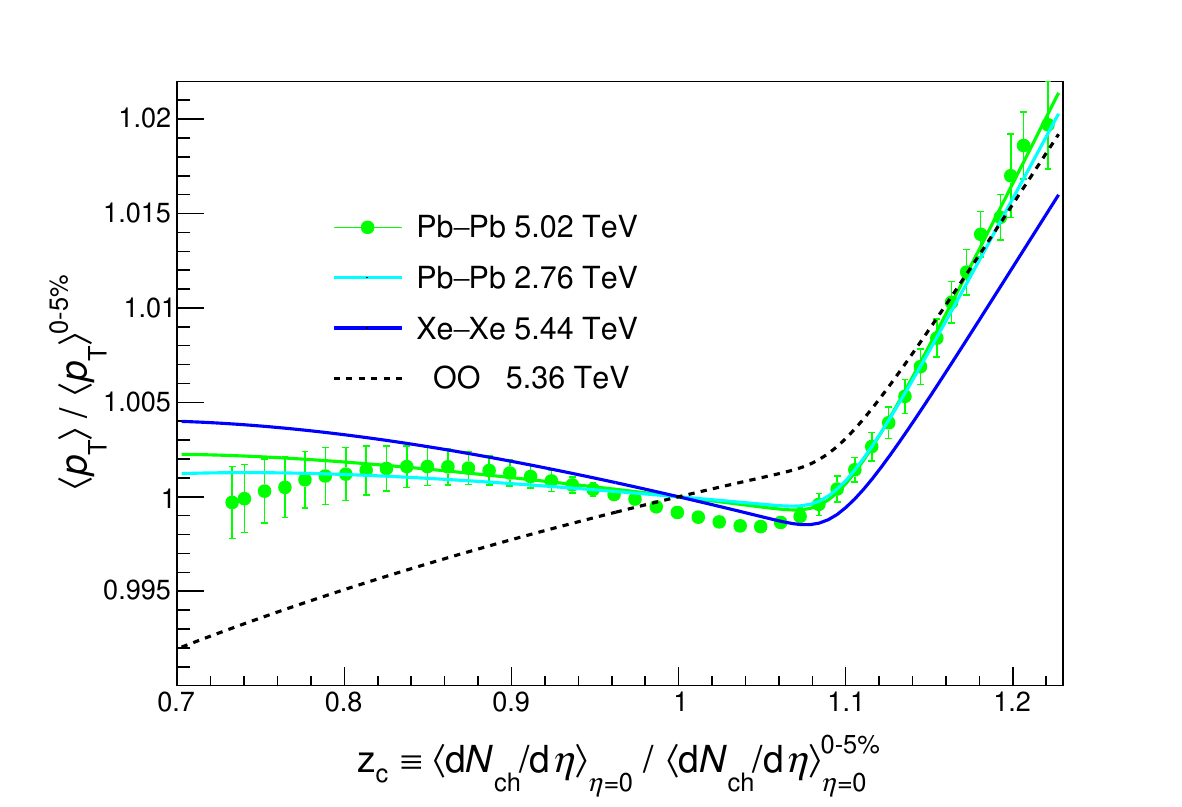}
\caption{Fit of charged particles normalized \mpt as a function of their normalized multiplicity density $z_c$ in \pbpb collisions at \enn $=$ 5.02~TeV~\cite{cms2401}. The predictions for \pbpb at \enn $=$ 2.76~TeV,  \xexe at \enn $=$ 5.44~TeV, and OO at $\enn=$5.36~TeV are shown too.} \label{nmeanpT}
\end{figure}

\noindent
longer than this medium~\cite{alic1}.

Next, Fig.~\ref{meanpT} shows the fit of the data on charged-particle \mpt versus \mch in \pbpb and \xexe collisions~\cite{alic2211}.  These data involve much more values of \mch 
than the \pt-spectra measurements~\cite{alic12,alic30}. The prediction for OO collisions is shown too. The sharp rise in model curves at very high \mch, related to factor $f_{uc}$ of Eq.~(\ref{eq:rho0}), is demonstrated in detail in Fig.~\ref{nmeanpT} with the \pbpb data~\cite{cms2401} at \enn$=$ 5.02~TeV for the normalaized \mpt and \mch to the corresponding values for 0-5\% centrality. The predictions for \pbpb at \enn $=$ 2.76~TeV,  \xexe at \enn $=$ 5.44~TeV, and OO at $\enn=$5.36~TeV are also shown. Overall, a good description of the data can be seen.

\subsection{\label{ssec2}Hadrons \pt-differential and averaged \vv}

It should be noted that the elliptic flow \vv data considered in present paper~\cite{alic10,alic21,alic22,alic23,cms4,cms5,alic24,alic28,alic1903,alic31,star3} were measured mostly by the two-particle cumulant method and partly by other similar methods that give compatible results within the data uncertainties.
This method assumes a minimum pseudo-rapidity gap $\Delta\eta$ between the two particles, meant to suppress the non-flow correlations, for which the largest available value in any dataset used was chosen.

\begin{figure}[ht]
\includegraphics[width=1.03\columnwidth]{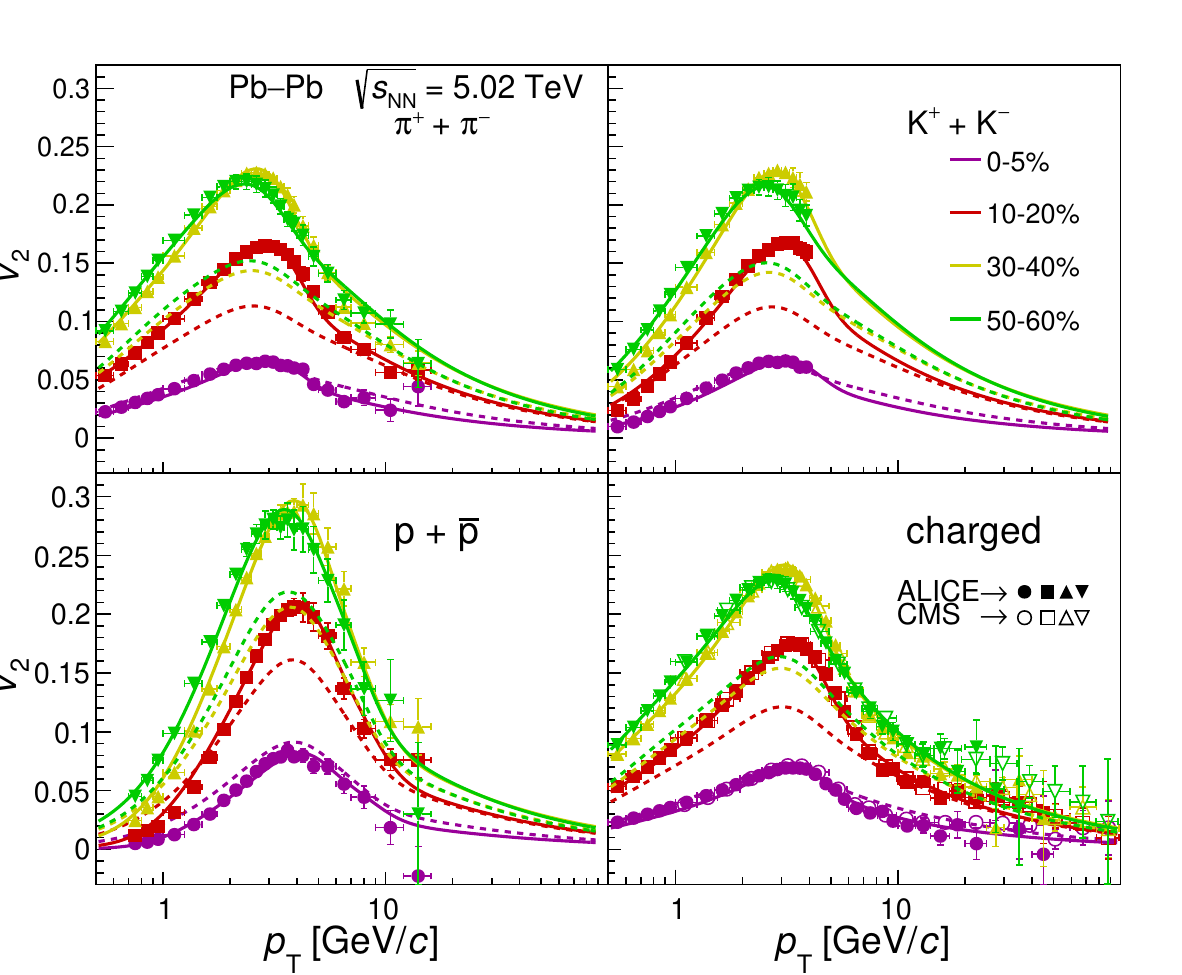}
\caption{Fit of pion, kaon, proton, and charged-particle \vv as a function of \pt in \pbpb collisions at \enn $=$ 5.02~TeV for four centrality intervals, measured by ALICE~\cite{alic22,alic23} and CMS~\cite{cms4} experiments. The dashed lines show the predictions for OO collisions at $\enn=$5.36~TeV.} \label{v2vspT502}
\end{figure}

\begin{figure}[ht]
\includegraphics[width=1.03\columnwidth]{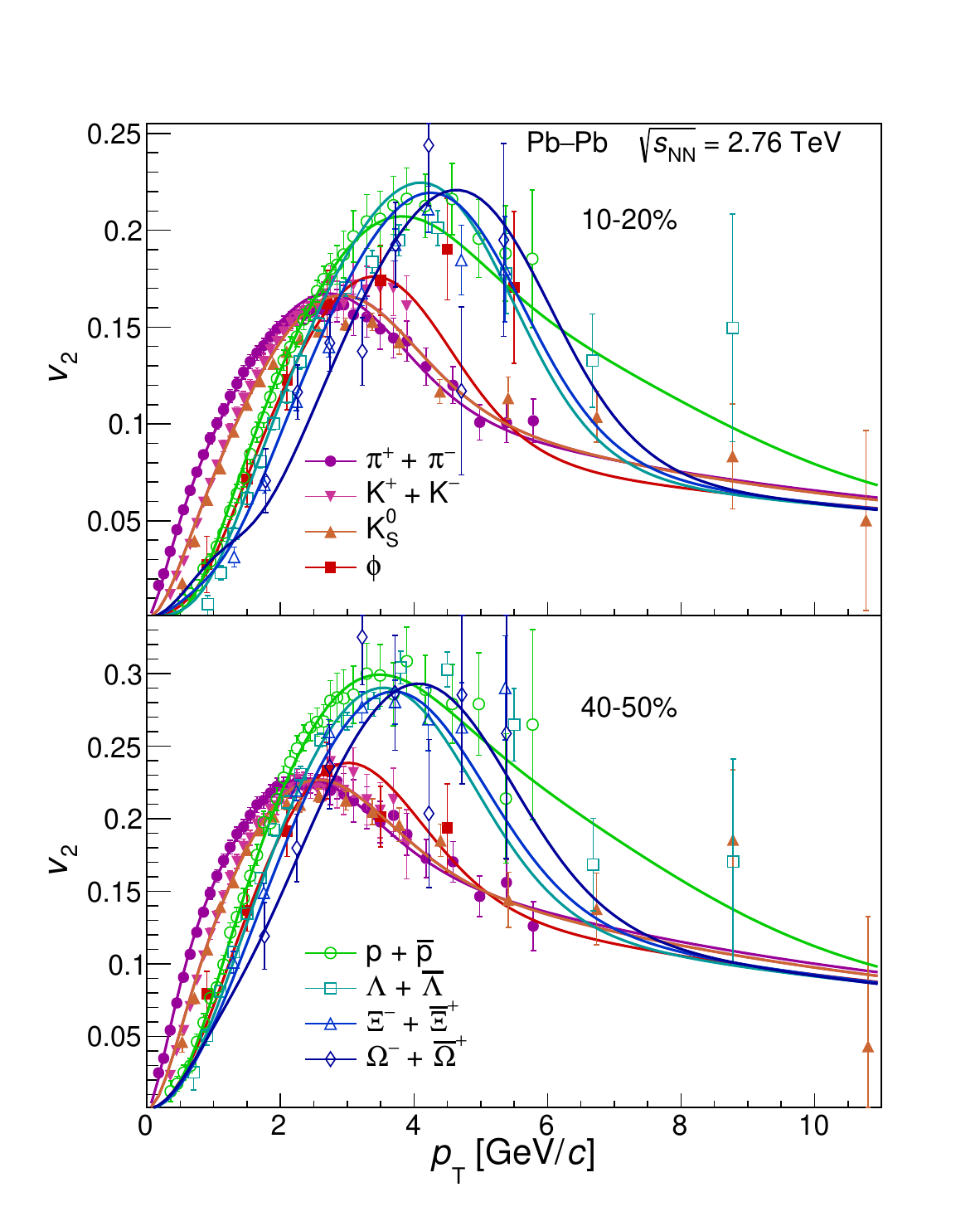}
\caption{Fit of \vv for different particles as a function of \pt in \pbpb collisions at \enn $=$ 2.76~TeV for centrality intervals 10-20\% and 40-50\%~\cite{alic21}.} \label{v2vspT276}
\end{figure}

\begin{figure}[H]
\includegraphics[width=1.0\columnwidth]{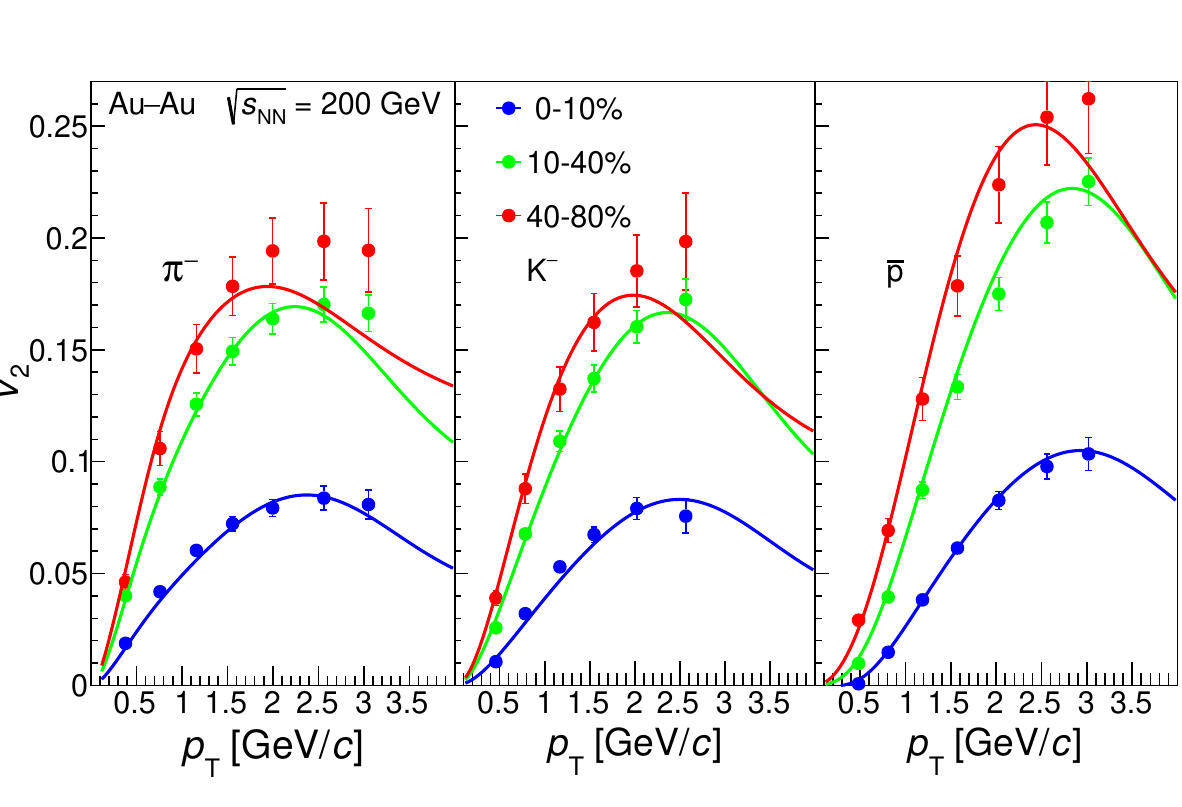}
\caption{Fit of \vv for \pim, \kam, and \pbar~ as a function of \pt in \auau collisions at \enn $=$ 200~GeV for three centrality intervals~\cite{star3}.} \label{v2vspT200}
\end{figure}

Figures~\ref{v2vspT502}, \ref{v2vspT276}, and \ref{v2vspT200} demonstrate the fits of  \pt-differential \vv for various hadrons and different centralities in \pbpb~\cite{alic21,alic22,alic23,cms4} and \auau~\cite{star3} collisions, as well as some features of such data~\cite{alic1,alic28}. A similarly good fit is also obtained for the \vv data in \xexe collisions at \enn $=$ 5.44~TeV~\cite{alic10,alic31}. The \mbox{\vv(\pt)} has a wide peak at $\pt\approx$ 1--7~\gevc whose height increases with the percentage of centrality up to $\sim$50\% and then decreases for more peripheral collisions. This is related to the collision geometry~\cite{alic28} and can be explained in the present model by the similar centrality dependence of \ecc, defined by Eqs.~(\ref{eq:ecc2}) and (\ref{eq:rxy}).
The peak is shifted to higher \pt for heavier particles. It is the well-known mass-ordering effect due to the radial flow~\cite{alic1,alic28}. Such a shift also occurs with increasing centrality (or \mch) due to the increasing radial flow velocity $\bett$ in  Eq.~(\ref{eq:bt}) caused by the decreasing exponent $k$ defined in Eq.~(\ref{eq:Tk}). Other known feature of \vv is the meson-baryon particle type grouping~\cite{alic1,alic28}, clearly visible in Fig.~\ref{v2vspT276}. This is described in present model by using different parametrizations for mesons and baryons in Eq.~(\ref{eq:p2}). 
Another effect is the increase of \vv with \enn~\cite{alic1}, which can be seen by comparing Figs.~\ref{v2vspT502} and \ref{v2vspT200}. It is mainly related to the increase of $\bett$ caused by the decreasing $k$ in Eq.~(\ref{eq:Tk}) with increasing \enn (or \mch). For the fit shown in Fig.~\ref{v2vspT200} the \pim, \kam, and \pbar~dataset was used since it includes more data

\noindent
points than the \pip, \kap, and \p~dataset~\cite{star3}. Note also that the dashed lines in Fig.~\ref{v2vspT502} show the predictions for OO collisions at $\enn=$5.36~TeV. 

It is important to note that the BW term of the present model (see  Eq.~(\ref{eq:v2})) gives, for most of considered hadrons, the main contribution to \vv in the lower part of its whole \pt-region up to the center of the peak. The Ts and Po terms, with the use of Eq.~(\ref{eq:v2tp}), describe \vv in the rest of this region.
However, for hadrons containing heavy quarks, for which the Ts term is zero and the BW term is suppressed by a factor $\sim\exp(-\mt /T)$, the Po contribution to \vv dominates everywhere (as for \pt-spectra~\cite{G21}).

\begin{figure}[ht]
\includegraphics[width=1.03\columnwidth]{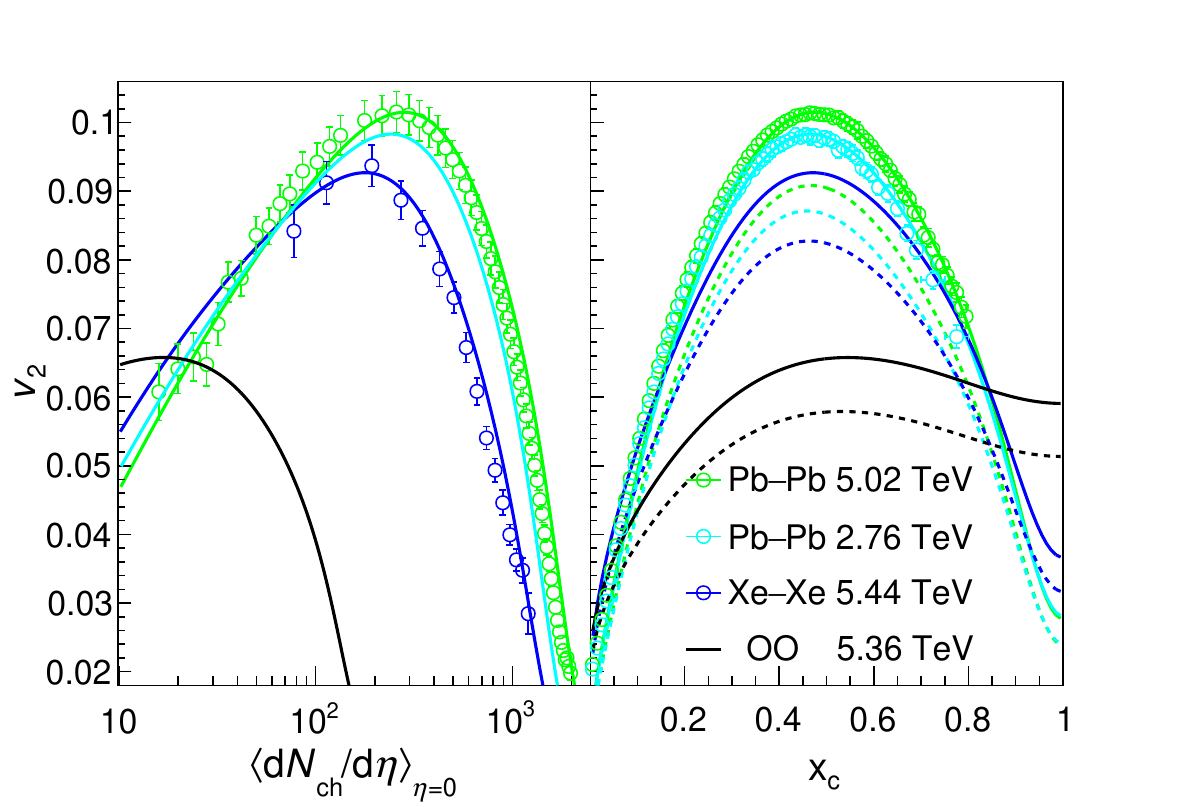}
\caption{Fit of charged-particle \vv, averaged in the \pt range $0.2<\pt<3.0$~\gevc, as a function of multiplicity density~\cite{alic1903} (left panel) and $x_c$~\cite{alic23} (right panel) in \pbpb collisions at \enn $=$ 5.02 and 2.76~TeV, and in \xexe collisions at \enn $=$ 5.44~TeV. The black lines show the predictions for OO collisions at $\enn=$5.36~TeV. The dashed lines correspond to \vv averaged in the \pt range $0<\pt<20$~\gevc.} \label{v2vsMxc}
\end{figure}

Finally, let us discuss the \pt-averaged \vv, defined via Eq.~(\ref{eq:v2N3}) with integration by \pt in the numerator and denominator.
Figure~\ref{v2vsMxc} shows the fit for the charged-particle \vv, averaged in the \pt range $0.2<\pt<3.0$~\gevc, as a function of \mch~\cite{alic1903} (left panel) and centrality fraction $x_c$ (defined in Sec.~\ref{sec3})~\cite{alic23} (right panel). Although both datasets were obtained by the two-particle cumulant method, they are independent measurements since Refs.~\cite{alic1903} and \cite{alic23} use $\Delta\eta > 1.4$ and $\Delta\eta>1$, respectively. Very good fits of the \pbpb and \xexe data can be seen in Fig.~\ref{v2vsMxc}. The black lines show the predictions for OO collisions at $\enn=$5.36~TeV. 

\noindent
For comparison, dashed lines are shown, corresponding to \vv averaged in the much larger \pt range $0<\pt<20$~\gevc. The left panel of Fig.~\ref{v2vsMxc} shows that the model describes well the different \mch (or centrality) dependences of \vv in \xexe and \pbpb collisions. This is due to the presence of the parameter $f_d$, related to the non-spherical form of $^{129}$Xe nucleus, and of the A-dependence in Eq.~(\ref{eq:rxy}). Also important is the smaller $\bett$ (caused by the larger $k$ from Eq.~(\ref{eq:Tk})) for collisions of smaller nuclei. These A-dependence and smaller $\bett$ are the main reasons for the lower values and flatter centrality (or \mch) dependence of the predicted \vv in OO collisions with respect to \pbpb collisions, shown in Figs.~\ref{v2vspT502} and~\ref{v2vsMxc}.

\subsection{\label{ssec3}Study of hadrons not used in the global fit}

To demonstrate the ability of BWTPM to describe the data for other hadrons, not used in the global fit, here I consider the ALICE measurements for the \pt-spectra of baryon resonances \Sirp and \Sirm \cite{alic32} and \lmbr \cite{alic33} in \pbpb collisions (the sum of the particle and antiparticle yields was measured in each case).
The data show that the yields of these short-lived resonances are suppressed in central collisions. This suppression can be quantified via the parameter $f_r$ from Eq.~(\ref{eq:Vs}), as for the \kstar~and \rhz resonances discussed above.
Since the data ~\cite{alic32,alic33} on baryon resonances are scarce, especially at low \pt, one can use for their parameter $f_r$ the value given in Table II. Then, to fit these data, there are left

\begin{figure}[H]
\includegraphics[width=1.03\columnwidth]{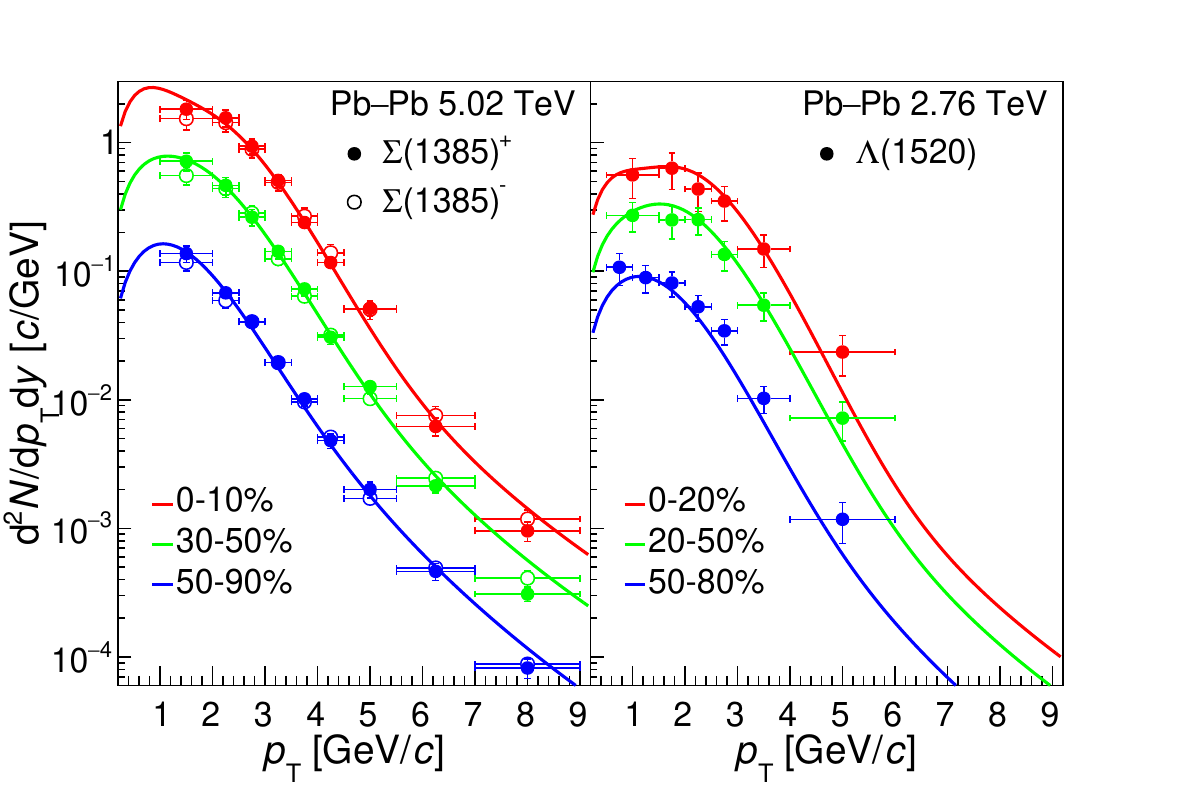}
\caption{Fit of \pt-spectra of \Sirs \cite{alic32} (left panel) and \lmbr \cite{alic33} (right panel) in \pbpb collisions at \enn $=$ 5.02 and 2.76~TeV, respectively, for three centrality classes.} \label{SigLam}
\end{figure}

\noindent
only three particle-specific free parameters: $f_{BW}, f_2$, and $h_n$. The combined fit of the data~\cite{alic32}, shown in the left panel of Fig.~\ref{SigLam}, gives $\chi^2/NDF = 6.8/51, f_{BW}=6.14, f_2=0.49$, and $h_n=1.25$. Note that the value of $h_n$ is close to the one given in Table I for hyperons. The \lmbr data~\cite{alic33} are very scarce at high \pt, so the fitting function is almost insensitive to the parameter $h_n$. The fit of these data, using fixed $h_n=1.25$ and two free parameters, gives $\chi^2/NDF = 5.4/17, f_{BW}=12.8, f_2=0.20$, and is displayed in the right panel of Fig.~\ref{SigLam}.
 
So, by fitting only few particle-specific parameters, BWTPM allows us to describe the data for new (not considered here) particles produced in AA collisions at different centralities and energies.

\vspace{2.0em}

\section{\label{sec7}Conclusions}

Thus, a more general and parametrized version of BWTPM~\cite{G21} is presented, which has much fewer free parameters, but successfully describes not only the midrapidity \pt-spectra, but also elliptic flow \vv of various hadrons with arbitrary \pt in AA collisions for any centrality at the LHC energies. It is shown that the model also works at much lower RHIC energies. For all considered particles, from pions to charmonia, BWTPM allows us to easily calculate the midrapidity \pt-spectra and \vv for new AA collisions with any A and centrality at high energies. Predictions are given for the OO collisions carried out recently at the LHC. 
However, for collisions of isobars and other deformed nuclei, shape-specific parameters should be introduced, such as the $f_d$ parameter used above for \xexe collisions.
To calculate the \pt-spectra or \vv for new particles, not considered in this study, one needs some data 
(e.g., \pt-spectrum for at least one centrality or \mpt versus centrality) to fix the particle-specific parameters, as shown for the strange baryon resonances in Sec.~\ref{ssec3}.

In the future, it is planned to employ BWTPM to describe the available data on the production of light nuclei (such as deuteron and helion) in AA collisions.

BWTPM can be efficiently used by experimenters to quickly check new data on hadron \pt-spectra and \vv.
Similarly, the simple parametrization for the charged-particle multiplicity density at midrapidity in AA or \pp collisions presented in Sec.~\ref{sec3} can also be used.

\section*{Acknowledgments}

I thank Jurgen Schukraft for valuable comments on the BWTPM parametrization.
I also thank the anonymous referee for important comments that led to further discussion of deformed nucleus collisions and BWTPM parameters as well as the addition of Sec.~\ref{ssec3}.

\section*{Data Availability}

The data supporting the findings of this article are openly available~\cite{alic4,alic5,alic6,alic536,phob,phen1,alic2308,
alic7,alic8,alic9,phob,T77,alic2211,cms2401,alic11,alic12,atl1,cms1,cms2,alic13,alic14,alic15,alic16,alic17,cms3,atl2,alic18, alic19,alic20,alic21,cms4,cms5,alic22,alic23,alic24,alic25,alic26,alic27,alic28,alic1903,alic10,alic29,alic30,alic31,phen2,
star2,star3,star4,brah,alic32,alic33}. 
 

\end{document}